\documentclass[epjST]{svjour}

\usepackage{amsmath}
\usepackage{amssymb}
\usepackage[utf8]{inputenc}
\usepackage{graphicx}
\usepackage{subfig}
\usepackage{placeins}

\usepackage{color}
\usepackage[backend=biber,natbib=true,sorting=none,giveninits=true,doi=false,url=false,maxbibnames=99,bibstyle=science,citestyle=numeric-comp]{biblatex}
\input{bib_style.cfg}
\newcommand{\ex}[1]{\text{e}^{#1}}
\newcommand{\beginsupplement}{%
        \setcounter{table}{0}
        \renewcommand{\thetable}{S\arabic{table}}%
        \setcounter{figure}{0}
        \renewcommand{\thefigure}{S\arabic{figure}}%
        \setcounter{section}{0}
        \renewcommand{\thesection}{S\arabic{section}}%
     }

\title{GranoEnv Paper Ideas}

\date{June 2020}

\begin{document}
\title{Anticipation-induced social tipping -- Can the environment be stabilised by social dynamics?}

\author{Paul Manuel M\"uller\inst{1,2,3}\fnmsep\thanks{\email{pmueller@pik-potsdam.de}} 
\and 
Jobst Heitzig\inst{1}
\and 
J\"{u}rgen Kurths\inst{1,4}
\and
 Kathy L\"{u}dge\inst{2}
 \and
Marc Wiedermann\inst{1}}

\institute{
FutureLab on Game Theory and Networks of Interacting Agents, Potsdam Institute for Climate Impact Research, Member of the Leibniz Association, PO Box 60 12 03, 14412 Potsdam, Germany \and 
Institute for Theoretical Physics, Technische Universit\"{a}t Berlin, Hardenbergstraße 36, 10623 Berlin \and
Department of Neurology, Charit\'{e} - Universit\"{a}tsmedizin Berlin, Charit\'{e}platz 1, Berlin 10117, Germany \and 
Centre for Analysis of Complex Systems, Sechenov First Moscow State Medical University, Moscow, Russia }

\abstract{In the past decades human activities caused global Earth system changes
, e.g., climate change or biodiversity loss. Simultaneously, these associated impacts have increased environmental awareness within societies across the globe, thereby leading to dynamical feedbacks between the social and natural Earth system. Contemporary modelling attempts of Earth system dynamics rarely incorporate such co-evolutions and interactions are mostly studied unidirectionally through direct or remembered past impacts. Acknowledging that societies have the additional capability for foresight, this work proposes a conceptual feedback model of socio-ecological co-evolution with the specific construct of anticipation acting as a mediator between the social and natural system. Our model reproduces results from previous sociological threshold models with bi-stability if one assumes a static environment. Once the environment changes in response to societal behaviour the system instead converges towards a globally stable, but not necessarily desired, attractor. Ultimately, we show that anticipation of future ecological states then leads to metastability of the system where desired states can persist for a long time.
We thereby demonstrate that foresight and anticipation form an important mechanism which, once its time horizon becomes large enough, fosters social tipping towards behaviour that can stabilise the environment and prevents potential socio-ecological collapse.
} 
\maketitle

\section{Introduction}

In the last decades humanities' impacts on the environment became increasingly more global in nature, more rapid, extensive, and also threatening for the social and environmental systems themselves \citep{Crutzen2002,Lenton2008,Steffen2011}. Scientists refer to this geological era as the \textit{Anthropocene} due to the impact humanity has and has had on the \textit{Earth system} \citep{Crutzen2002, Steffen2011}. In the Anthropocene global challenges like climate change and resource scarcity need to be addressed and interventions are required in order to build a sustainable society which is able to overcome these challenges \citep{Steffen2011a}.

To improve the understanding of Earth system dynamics in the Anthropocene, it is crucial to model both the social and the ecological systems as coupled together due to the interconnections and bidirectional feedback processes between humanity and the environment \citep{ Schellnhuber1998, Palmer2014,Muller-hansen2017,Donges2017,Donges2018a,Donges2018,Bury2019}, given rise to the novel field of \textit{World-Earth modeling} \citep{Donges2018}. Examples of such models include network models of individual resource use and social adaptation  \citep{Wiedermann2015,Barfuss2017} combined with effects of governance and taxation \citep{Geier2019}. Other models investigate land-use and social learning \citep{Muller-Hansen2019} or land-use and management \citep{Brown2019}. There are also models which incorporate not only a social and ecological system but additionally include an economic branch \citep{Nitzbon2017}.

One important feature of the World-Earth system are \textit{tipping elements}. Tipping elements change rapidly and qualitatively after surpassing a critical threshold, the so-called \textit{tipping point} \citep{Lenton2008}.  In the past many natural tipping elements, like the Greenland-ice sheet which can either persist or melt-off completely due to melt-elevation feedback or the Amazon, which does either exist in today's rainforest state or transition into a Savannah-like state, have been identified and their risk of tipping even within the 2$^\circ$ C target of the Paris agreement has been investigated \citep{Lenton2008, Schellnhuber2016}. 

On the other hand, \textit{social tipping} and its potential to transform the society so that the aforementioned (unwanted) natural tipping may be prevented came into research focus recently \citep{Otto2020, Milkoreit2018, DavidTabara2018, Wiedermann2020}. Social tipping can for example be observed when a determined minority grows larger than a critical mass and then can be able to overturn the behaviour or conventions of the majority \citep{Xie2011,Centola2018}. Granovetter's threshold model is an early model formulating and explaining this phenomenon on the example of riots \citep{GranThres}. For the binary decision of not participating or participating in a riot it is assumed that every individual needs to observe a certain percentage of the population 
participating in the riot before doing so themselves. This percentage 
is called their \textit{(population-pertaining) threshold}. From the distribution of the thresholds one could then calculate the equilibrium number of participants in the riot. However, real threshold distributions are hard to evaluate and reasonable guesses for the distributions lead to everyone or no-one participating in the riot \citep{Wiedermann2020}. Two possible extensions to tackle these issues are the introduction of determined minorities and the explanation of a person's apparent population-pertaining threshold in terms of the person's position in the social network and a uniform \textit{neighbour-pertaining threshold} \citep{Wiedermann2020, Watts2002}. This latter type of threshold determines what percentage of a person's acquaintances need to participate for the person to participate itself. The wide distribution of people's population-pertaining thresholds then arises because an emerging cascade of activity reaches different locations in the network at different stages of the riot.

 While there is already quite a consolidated understanding of the dynamics within single tipping elements \citep{Feudel2018,Kaszas2019}, especially from the natural realm \citep{Boers2017, Levermann2016}, it remains an important task of ongoing research to study interactions between different elements, specifically across the climate and the social system \cite{Donges2018a,Winkelmann2019}, since exemplary studies have shown that such interactions can give rise to substantial changes of such systems on the macro-scale~\citep{ricke2014natural, Beckage2018}. Recent work has investigated also the interaction of  tipping elements in natural \citep{Kriegler2009,Wunderling2020, Klose2020,Kroenke2020,Wunderling2020b}, social \citep{Otto2020},  socio-economic \citep{Rocha2018} or socio-ecological systems \citep{Brummitt2015}.
However, most of the aforementioned studies have either studied tipping elements by using identical prototypical normal-forms for each element \citep{Wunderling2020,Klose2020,Kroenke2020,Wunderling2020b,Brummitt2015} or by focusing on cascading effects and interactions within either only natural or social subsystems \citep{Kriegler2009,Wunderling2020,Klose2020,Kroenke2020,Wunderling2020b,Otto2020} and thus without explicitly accounting for their potential cross-system interactions. 

In order to bridge this gap, the present work investigates potentially arising dynamics of coupled tipping elements across the natural and social system and proposes a conceptual, low-dimensional model consisting of 
a stylised environment being described by one macroscopic variable which we call the \textit{pollution} and a social network of individuals which change their behaviour, i.e., either their contribution to the pollution or their active decision to not contribute to it, according to change probability rates based on three threshold processes:
\begin{enumerate}
    \item \textit{Direct environmental impacts} like droughts, heatwaves or air pollution 
can threaten human society \citep{Patz2005,Brook2010,McMichael2012}.  An adaptation to and mitigation of these impacts is necessary in a changing
Earth system \citep{McMichael2011}.
\item 
\textit{Social contagion} which is recognised as the 
spread of behavioural patterns among
many individuals in an interacting society \citep{Centola2007}, e.g.\ social
movements \citep{Gould1991,Gould1993,Zhao1998} or the spread of social norms \citep{Centola2005}.
\item \textit{Anticipated environmental impacts} are rarely discussed in World-Earth models and the social system are mostly coupled through precedent or contemporary effects to the environment. However, societies having the capability of anticipation and foresight might alter the social behavioural patterns significantly. For example, the global climate changing to the
worse, like in a  ``Hothouse Earth'' scenario \citep{Steffen2018}, could
spark and strengthen climate movements 
 
with the goal to
prevent consequential environmental impacts \citep{Hagedorn2019}. Especially,
information and the anticipatory time horizon up to which an individual projects such impacts
can have significant influence on the behaviour of individuals and the macroscopic dynamics
 \citep{Bury2019,Pahl2014}. 
\end{enumerate}

In particular, 
we 
incorporate the social system via an 
agent-based model (ABM)
similar to Ref.~\cite{Wiedermann2020}, since ABMs have 
proved as a promising tool to model social behaviour and decision making  \citep{Epstein1999,Bonabeau2002,Macy2002}.  In ABMs individuals, called \textit{agents}, 
adjust their behaviour according to certain deterministic or stochastic rules after assessing their individual situation \citep{Bonabeau2002}. One of the strengths of ABMs lies in the microscopic details which arise 
from the simulation of individuals and can induce emergent macroscopic phenomena 
\citep{Page2015}. Within ABMs non-linear features, such as local heterogeneous structures,  can be observed and analysed to help understanding the impact of disturbances or causes of unexpected outcomes \citep{Muller-hansen2017}. However, ABMs come with significant computational costs and are hard to analyse formally. Therefore, 
we 
also apply a mean-field approximation of the microscopic behaviour of the network of interacting agents.

vironment.  

The model simulations presented in this work reveal that a direct coupling between a dynamic environment and social dynamics leads to one globally stable attractor. However, additionally accounting explicitly for the effects of anticipation induces metastability where additional environmental states that are not within or close to that stable attractor persist over a long time. 

The remainder of this paper is structured as follows. First, in Section\;\ref{sec:model} the micro-level 
socio-ecological model itself is introduced and the implementation of the three aforementioned processes is explained. Second, 
a 
mean-field approximation of the model is given in Section\;\ref{sec:approx}. 
Third, we give an overview of the model behaviour, first 
for a static and then a dynamic environment, and 
compare it to the mean-field approximation in Section\;\ref{sec:case1_2}.
In Section\;\ref{sec:bif_theta} one can find the main results of the paper, 
metastability for high anticipation times resulting in a stabilisation of potentially unpolluted environments and thus social tipping. 
We close this paper with a discussion of the results, possible connections to real-world parameters and a short outlook in Section\;\ref{sec:disc}.

\section{Model}
\label{sec:model}

\begin{figure}[tb]
    \centering
    \hspace{1.8cm}
    \includegraphics[width=.7\textwidth]{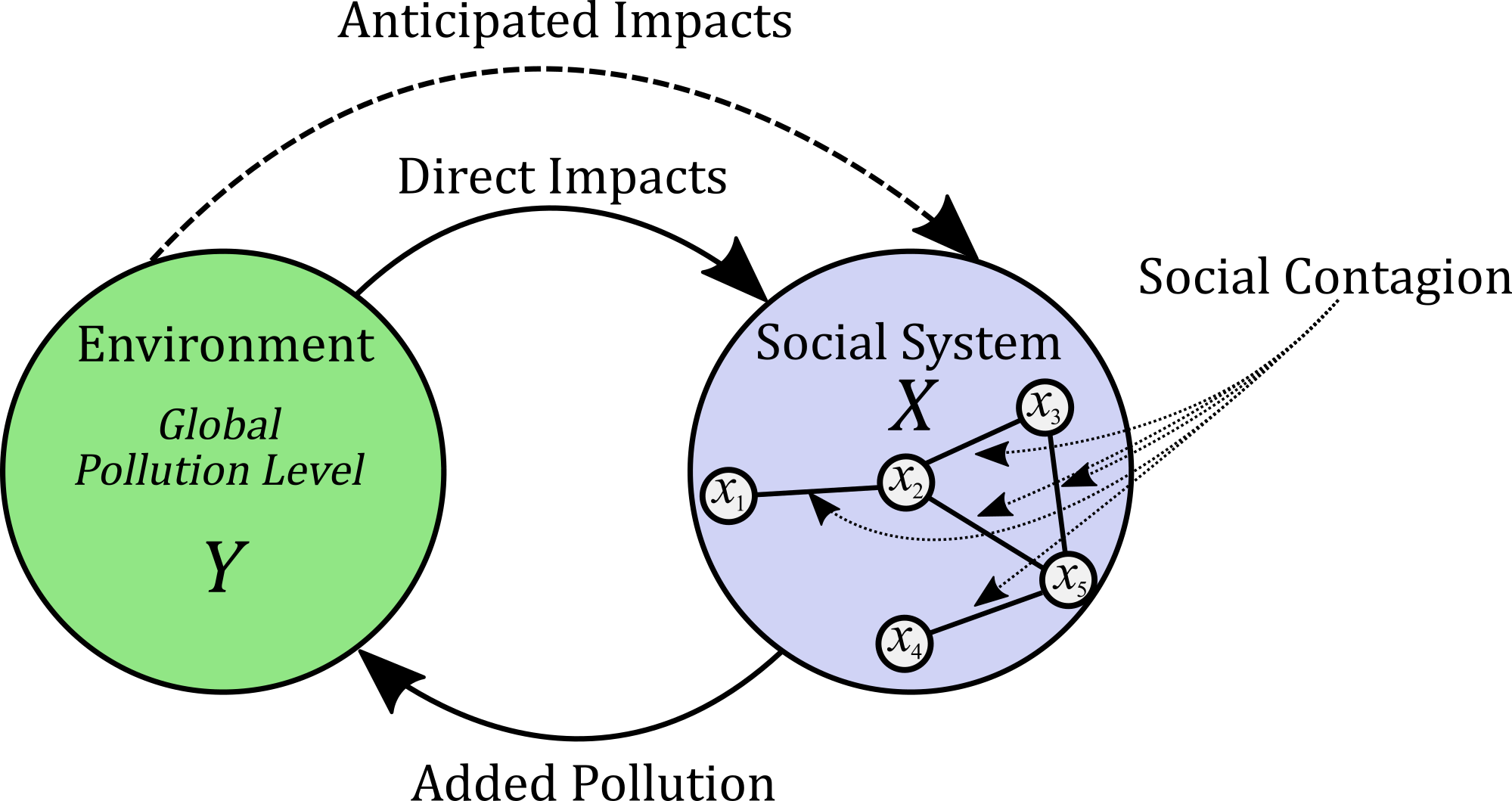}
    \caption[Schematics for the model setup]{Schematics for the model setup. The environment is represented by just 
    one variable, the global pollution level $Y$, which changes proportional to its own value and the social systems mean state, the share of polluting agents $X$. The social system consists of individual agents which are either adding to the pollution or not. They change their binary state $x_i$ depending on the current pollution level (direct impacts), the pollution level they anticipate in the future (anticipated impacts) and the behaviour of their direct neighbours in the network (social contagion). }
    \label{fig:GranoEnv}
\end{figure}

To illustrate the influence of the environment on social activation and vice versa we formulate a conceptual network model which is tuneable  from a  mainly social dynamic version similar to Granovetter's threshold model \citep{GranThres} as in Ref.~\citep{Wiedermann2020} to a coupled socio-ecological agent-based model. A  focus in the model is 
the effect of anticipation. In Fig.\;\ref{fig:GranoEnv} we provide a schematics of  the model's setup which illustrates the different 
interaction mechanisms between the environment and the social system, explained in detail below.

First, we define agent $i$'s pollution state 
as
\begin{align}
x_i = \begin{cases}
    1,\;\text{for }\textit{polluting} \text{ (i.e., not mitigating pollution) agents}\\
    0,\;\text{for }\textit{non-polluting}\text{ (i.e., mitigating pollution) agents}
    \end{cases}.\label{eq:states}
\end{align}
Agents are able to switch from both states to the other and especially the direction from polluting to actively non-polluting is of importance if one wants to guide the system towards an unpolluted state.
The mean share of polluting agents will be denoted 
$X= \frac{1}{N}\sum_i x_i$ where $N$ is the number of agents. We 
use this macroscopic variable $X$ to couple the social system to the environment which is represented by only one macroscopic variable, the pollution $Y$. 

\subsection{Ecological Dynamics}

As greenhouse gases, especially CO$_2$, are 
the major driving force
of anthropogenic climate change \citep{Hansen1981,Stott2000}, we motivate our environmental module with their fundamental dynamics. Carbon added to the atmosphere, for example by burning fossil fuel, does not stay in the atmosphere but is constantly removed
through natural carbon sinks like terrestrial ecosystems or the ocean, with the latter being the major sink in the Earth system \citep{Cox2000,Sabine2004}. For simplicity, the carbon decay is modelled by a linear
differential equation with a single average lifetime $\tau$ \citep{Moore1994,Cox2000,Kellie-Smith2011}. Additionally, pollution is added proportionally to the share of polluting agents $X$, 
\begin{align}
 \dot Y = \frac{1}{\tau} (X-Y).\label{eq:Ydot}
\end{align}
In particular, $Y$ is measured in units of \textit{ maximal equilibrium pollution}, so that if all $N$ agents pollute consistently ($X=1$), then $Y\to 1$.
The 
average lifetime of 
pollution,  $\tau > 0$, allows to scale the system to a quasi static environment by letting 
$\tau\rightarrow\infty$. However, a reasonable 
estimate for  an average lifetime 
is
$\tau\approx50$\,years \citep{Moore1994,Sabine2004}. We are aware that this implementation of the environment is very conceptual and that there are also comparatively more advanced models of the carbon cycle \citep{Cox2000,Firedlingstein2006, Sitch2008,Anderies2013}. 
However, for the qualitative behaviour of our model only some form of natural decay is necessary, so the environmental dynamics from Eq.\;(\ref{eq:Ydot}) suffices.

\subsection{Social Dynamics} 

The social dynamics of the share of polluting agents $X$ arise from 
the dynamics of the single agents' states ${\bf x} = (x_i)_{i=1}^N$. 
The agents change their states with the change probability rates $p_i^-({\bf x},Y)$ from polluting to non-polluting ($x_i=1\rightarrow x_i=0$) and $p_i^+({\bf x},Y)$ from non-polluting to polluting ($x_i=0\rightarrow x_i=1$): 
\begin{align}
p_i^-({\bf x},Y) &= \alpha\cdot p_\text{dir}^-(Y) + \beta \cdot p_{i\text{,soc}}^-({\bf x})\cdot p_\text{ant}^-(X,Y),\label{eq:stylizedpp}\\
p_i^+{\bf x},Y) & =\alpha\cdot p_\text{dir}^+(Y) + \beta \cdot p_{i\text{,soc}}^+({\bf x})\cdot p_\text{ant}^+(X,Y).\label{eq:stylizedpm}
\end{align}
There are three contributions to the change probability rates which we will motivate and explain in detail in the following paragraphs: those due to direct environmental impacts, $p_\text{dir}^\pm(Y)$, due to social contagion, $p_{i,\text{soc}}^\pm({\bf x})$, and due to anticipated environmental impacts, $p_\text{ant}^\pm(X,Y)$. The first contribution is scaled by a parameter $\alpha \ge 0$ which we call the \textit{vulnerability}, since one could interpret it as how susceptible an agent is to immediate changes in their environment. The second and third contributions are coupled together and are scaled by a parameter $\beta \ge 0$ which we call the \textit{farsightedness} since it weights the anticipation effect. This coupling can be understood as social contagion leading to a spread of the action due to anticipated environmental impacts. 
Note that among these factors, only $p_{i,\text{soc}}^\pm({\bf x})$ depends on the microstate $\bf x$ while the others only depend on the macrostate $(X,Y)$.

\paragraph{Direct environmental impacts.}
Direct environmental impacts like droughts or storms have triggered sudden societal changes in the past, including the onset of migratory patterns or even the collapse of whole nations \citep{Cullen2000,DeMenocal2001}. In our model we 
link the behaviour of agents to the current pollution level $Y$ through a threshold function. If the pollution is above the \textit{pollution threshold} parameter $\gamma\in[0,1]$, a polluting agent has a higher probability to become non-polluting and vice versa,
\begin{align}
p_\text{dir}^-(Y) &= H\left(Y-\gamma\right),\\
p_\text{dir}^+(Y) &= 1-H\left(Y-\gamma\right),
\end{align}
where $H$ is the Heaviside 
step function.

\paragraph{Social contagion.}
The embedding of individuals in a social network and the observation of their immediate surroundings has been
proven to have a large influence on people's behaviour \citep{Centola2005,Centola2007,Christakis2013, Melnik2013}. Recently proposed network models allow to incorporate and explain such dynamics in the context of opinion formation \citep{Holme2006}, behaviour \citep{Christakis2013,Centola2010} and  information spread \citep{Monsted2017} or social movements and collective action \citep{Melnik2013}.
Granovetter's threshold model \citep{GranThres} and recent network based extensions  \citep{House2011,Singh2013,Karsai2016,Wiedermann2020} describe such influence by other individuals or direct neighbours, also known as \textit{social contagion} \citep{Burt1987}. From Refs.~\cite{Watts2002,Wiedermann2020} we take the idea to define a microscopic \textit{neighbor-pertaining 
threshold} $\chi\in[0,1]$ which influences the likelihood of the agents' behavioral change. 
If the share of polluting
neighbours in the fixed social network $G$, $\frac{X_i}{k_i}$, is lower than the social threshold $\chi$,  a polluting agent becomes more likely to convert to being non-polluting and vice versa. 
In this, $X_i = |\{j:(i,j)\in E,~x_j=1\}|$ gives the number of polluting neighbours, $k_i = |\{j:(i,j)\in E\}|$ is $i$'s degree (number of neighbours),
and $E$ is the fixed set of network edges. 
\begin{align}
p_{i,\text{soc}}^-({\bf x}) &= 1-H\left(\frac{X_i}{k_i}-\chi \right),\label{eq:socp+}\\
p_{i,\text{soc}}^+({\bf x}) &= H\left(\frac{X_i}{k_i}-\chi \right).\label{eq:socp-}
\end{align}

\paragraph{Anticipated environmental impacts.}
The anticipation of potential environmental catastrophes might
spark, strengthen, and justify climate movements like \textit{Fridays for Future} \citep{Hagedorn2019}. Specifically, the anticipation time, i.e., how far one extrapolates the trajectory of the Earth system into the future, can have a big impact on the social urge to act \citep{Pahl2014,Bury2019}. We assume the simplest possible form of anticipation, a linear extrapolation, and therefore 
define the anticipated pollution  as
\begin{align}
Y_\text{ant}=Y+\theta\dot{Y}\,,
\end{align}
where $\theta > 0$ is the \textit{anticipation time}. 
If the anticipated pollution is 
above the pollution threshold $\gamma$, a polluting agent becomes more likely to get non-polluting.
Likewise, if it is below, a non-polluting agent becomes more likely to get polluting 
\begin{align}
p_\text{ant}^-(X,Y) &= H\left(Y + \theta\dot{Y} - \gamma\right)
= H\left(Y + \frac{\theta}{\tau} (X-Y) - \gamma\right),\\
p_\text{ant}^+(X,Y) &= 1 - H\left(Y + \theta\dot{Y} - \gamma\right)
= 1 - H\left(Y + \frac{\theta}{\tau} (X-Y) - \gamma\right).
\end{align}
Plugging these factors 
into Eq.\;(\ref{eq:stylizedpp}, \ref{eq:stylizedpm}) yields the full change probability rates,
\begin{align}
 p^-_i({\bf x},Y) &=
    \alpha H(Y - \gamma)
    + \beta \left[1 - H(X_i - \chi k_i)\right]\cdot H\left(Y + \frac{\theta}{\tau} (X-Y) - \gamma\right), \\
 p^+_i({\bf x},Y) &=
    \alpha \left[1 - H(Y - \gamma)\right] 
    + \beta H(X_i - \chi k_i)\cdot\left[1 - H\left(Y + \frac{\theta}{\tau} (X-Y) - \gamma\right)\right],
\end{align}
which together with Eq.\;(\ref{eq:Ydot}) fully define our system for a given network structure.

\subsection{Mean-Field Approximation}
\label{sec:approx}

Applying a mean-field approximation and therefore getting rid of the explicit network dependence allows us to 
get important insights into the systems dominant dynamics. We therefore assume that the network is large, densely connected, and has a small diameter, such that changes spread fast through the network and thus, most of the time, the share of polluting neighbours is approximately equal for all nodes. Consequently, the latter can  be approximated well by the overall 
share of polluting agents,
\begin{align}
\frac{X_i}{k_i} \approx X.
\end{align}
Plugging this into Eq.\;(\ref{eq:socp+},\ref{eq:socp-}) cancels out the network dependence and all agents have approximately the same change probability rates,
\begin{align}
 p^-({\bf x},Y) &\approx\tilde p^-(X,Y)
 =  
    \alpha H(Y- \gamma) 
    + \beta \left[1 - H(X - \chi)\right]\cdot H\left(Y +
    \theta \dot Y
    - \gamma\right), \\
 p^+({\bf x},Y) &\approx \tilde p^+(X,Y) =  
    \alpha \left[1 - H(Y-\gamma)\right] 
    + \beta H(X - \chi)\left[1 - H\left(Y + \theta \dot Y
    - \gamma\right)\right].
\end{align}
Since the network is also assumed to be very large, the actual stochastic change in $X$ can be approximated by its expectation value and
we obtain finally the deterministic differential equations
\begin{align}
    \dot X &= (1 - X ) \tilde p^+(X,Y) - X \tilde p^-(X,Y),  \label{eq:MF}\\
    \dot Y &= \frac{1}{\tau} (X - Y).\nonumber
\end{align}
Through this approximation we reduced the number of equations from $2N+1$ to only two (however non-smooth) differential equations which can 
be solved numerically. 
In addition, a piecewise analytical solution is possible, since the change rates $p^\pm$ are constant except when $X$ meets the threshold $\chi$ or when $Y$ or $Y_\text{ant}$ meets the threshold $\gamma$. For constant $p^\pm$, the differential equations become linear in $X$ and $Y$ and can be solved through an exponential ansatz and the method of varying constants (see S1). However, due to the non-smoothness it is not possible to find a full solution,  only such a piece-wise solution.
For the numerical implementation we used the piece-wise calculated equations for $X$ and $Y$ and probed through integrating for discrete time steps  which piece-wise solution has to be used. 
\footnote{Implementations of the mean-field model and the agent-based simulations are available online at \url{https://github.com/PaulMMueller/GranoEnv}.} 

\section{Results}

In the first part of this section, we 
compare the dynamics of the microscopic model to the dynamics of the above introduced mean-field approximation. First, the case of a static environment is studied, i.e., $\tau\rightarrow\infty$, to show that our model displays a bistable behaviour 
also found in other works on social tipping 
\citep{Wiedermann2020,Singh2013}. Second, we study the case of a dynamic environment 
and we investigate the attractors of this feedback system. 
Subsequently, we 
use the mean-field approximation and identify  multi-stability in the regime of long anticipation times.

\subsection{Comparing the Microscopic Model to the Mean-Field Approximation}
\label{sec:case1_2}

\begin{figure}
\includegraphics[width=\textwidth]{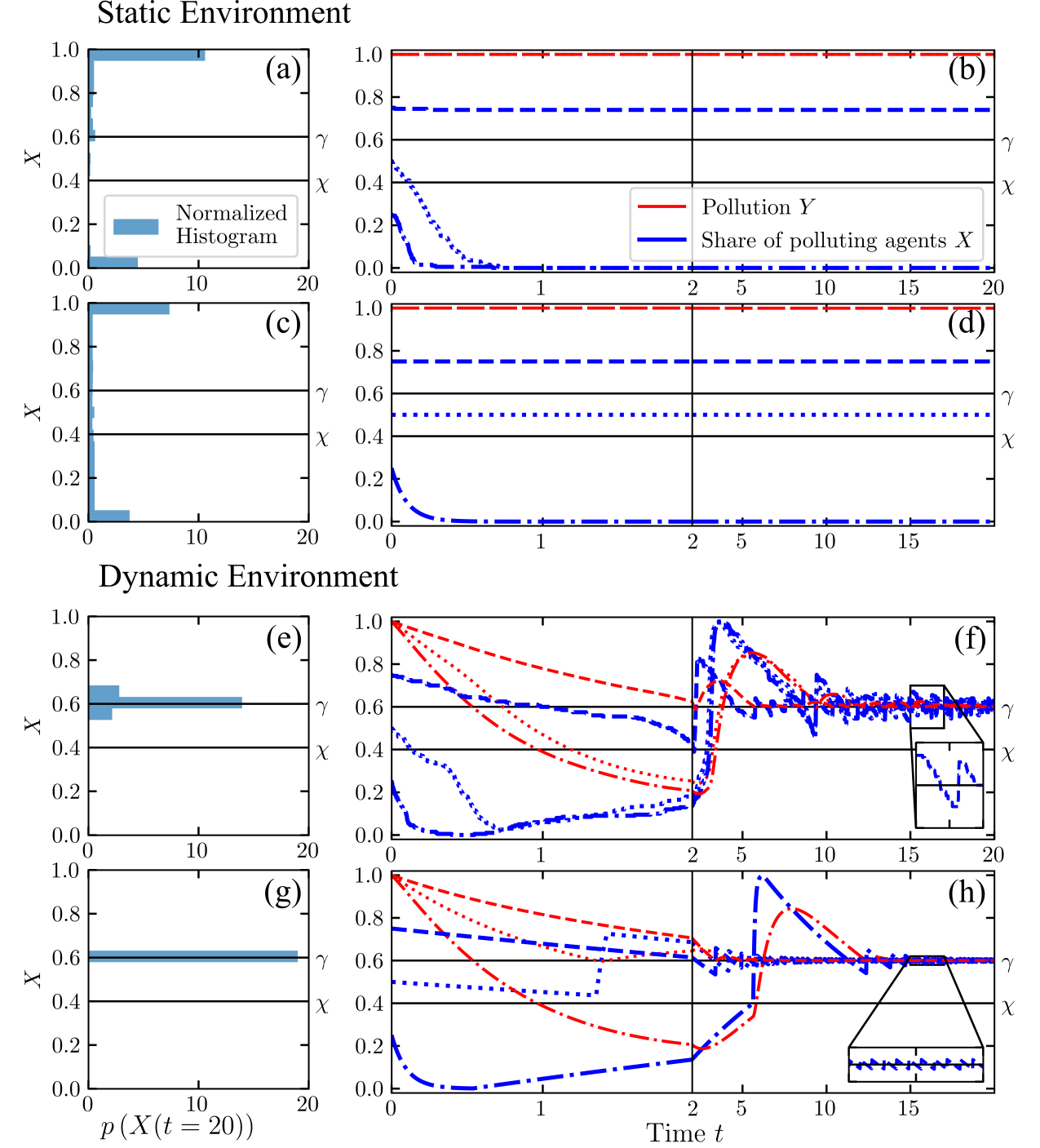}
\caption[Comparison of the microscopic model with the mean-field approximation for a static and a dynamical environment.]{Comparison of the microscopic model (panels a,b,e,f) to the mean-field approximation (c,d,g,h). \textbf{(a,b,e,f):} static environment with parameters $\gamma=0.6$, $\chi = 0.4 $, $\alpha = 0 $, $\beta = 10$, $\tau = 10^6$, $\theta = 1$, and three different initial conditions (dashed, dotted, dash-dotted). In (a,c)  the distribution of trajectories $p\left(X(t=20)\right)$ with  $X(t=20)$ is illustrated through a histogram. Generally, two main attractors at $X=0$ and $X=1$ can be seen. Panels (b,d) show trajectories with $Y(t=0)=1$ and three different $X(t=0)$ for the microscopic simulations and the mean-field approximation respectively. The black lines mark the thresholds $\gamma$ and $\chi$.
\textbf{(c,d,g,h):}
dynamical environment with $\tau=1$, a direct coupling of the social system to the environment, $\alpha=0.1$, and other parameters as before. There is only one main attractor at $X(t=20)\approx\gamma$ for both the microscopic simulations (e,f) and the mean-field approximation (g,h). Deviations can be found in the transient behaviour  ($t<2$) as seen in the trajectories shown in (f,h), while the long term behaviour is similar. In the insets oscillations around the pollution threshold $\gamma$ are shown for one trajectory ($X(t=0)=0.75,\,Y(t=0)=1$).}
\label{fig:case1_2}
\end{figure}

We compare the mean-field approximation with microscopic model simulations for Erd{\H{o}}s--R{\'{e}}nyi networks with $N=200$ nodes and an average degree of $\bar{k}=10$. 
Calculations are robust against larger network size so that finite size effects are negligible in the presented results (not shown).
In Fig.\;\ref{fig:case1_2}\;(a--d), results for a static environment ($\tau = 10^6$) are displayed. Additionally, the social system does not directly depend on the environment ($\alpha=0$) but only through the anticipated impacts ($\beta=10$). These parameter choices ensure that the dynamics is dominantly driven by social contagion, see Eq.\;(\ref{eq:stylizedpp},\ref{eq:stylizedpm}), so that we can compare our model in this boundary case against other studies \citep{Wiedermann2020,Singh2013}.
The microscopic results are shown in Fig.\;\ref{fig:case1_2}\;(a,b) and the mean-field approximation results in Fig.\;\ref{fig:case1_2}\;(c,d).  In Fig.\;\ref{fig:case1_2}\;(a,c) the distribution  of the long-term share of polluters, $p(X(t=20))$, is shown as a histogram. In both cases we observe two prominent maxima at $X=0$ and $X=1$ which compares well to previous network-based threshold models that are showing bistability as well \citep{Singh2013,Wiedermann2020}. The exemplary trajectories in Fig.\;\ref{fig:case1_2}\;(b,d)  show that the trajectories approach $X=0$ if $Y\geqslant\gamma$ and $X<\chi$. In the microscopic case, Fig.\;\ref{fig:case1_2}\;(b), trajectories with $X(t=0)$ slightly above $\chi$ can approach the attractor at $X=0$ even though the social system would be above the social threshold. This can be explained through the local heterogeneity of the network leading to some nodes in an environment with $\frac{X_i}{k_i}<\chi$ even though $X\geqslant\chi$. These nodes can then start a cascade eventually leading to all nodes  approaching the same state. In the mean-field approximation, Fig.\;\ref{fig:case1_2}\;(d), such a behaviour can not be observed as $X<\chi$ and $Y+\theta\dot Y\geqslant\gamma$ results in all change rates becoming zero.

In Fig.\;\ref{fig:case1_2}\;(e--h) we introduce a dynamic environment ($\tau =1$) and couple the social system directly to the environment ($\alpha=0.1$). The system again reaches its attractor before $t=20$, and in Fig.\;\ref{fig:case1_2}\;(e,g) the distribution of $X(t=20)$ is displayed. In contrast to the static environment case, it peaks around  $X\approx Y\approx \gamma$ for both the microscopic simulations Fig.\;\ref{fig:case1_2}\;(e) and the mean-field calculations Fig.\;\ref{fig:case1_2}\;(g). The existence of only one main attractor around $X\approx Y\approx \gamma$ is emphasized through the trajectories in Fig.\;\ref{fig:case1_2}\;(f,h) which in the long-term show the same behaviour for microscopic and mean-field simulations. However, in the transients $(t<2)$ the microscopic simulations still differ from the mean-field solution due to the heterogeneity of the underlying network. 
The coupling to a dynamic environment, Fig.\;\ref{fig:case1_2}\,(e--h), leads to a loss
of the multistability of the pure social model shown in Fig.\;\ref{fig:case1_2}\;(a--d)  which can be interpreted as a form of ``damping''. Additionally, one can observe that $|X-Y|$ becomes very small which should be expected since due to 
Eq.\;(\ref{eq:Ydot}) the pollution  always follows the value of $X$. However, $X$ and $Y$ do not completely converge 
but oscillate around the pollution threshold $\gamma$ which is illustrated in the insets in Fig.\;\ref{fig:case1_2}\;(f,h) for one $X(t)$ ($X(t=0)=0.75$). 
During each period of this oscillation, both $Y$ and $Y_\text{ant}$ cross the threshold $\gamma$ twice, so the trajectory is continuous but not smooth.
Due to the stochasticity of the microscopic simulations the oscillations in Fig.\;\ref{fig:case1_2}\;(f) are less regular than the oscillations in the mean-field approximation, see the inset in Fig.\;\ref{fig:case1_2}\;(h).

We saw that the mean-field approximation is not accurate in the transient behaviour, see Fig.\;\ref{fig:case1_2}\;(f,h) for $t<2$, but generally provides comparable results in the long-term behaviour. Therefore, it qualifies for a qualitative assessment of the model dynamics which we will use in the next section to analyse the trajectories for different anticipation times $\theta$.

\subsection{Metastability for High Anticipation Time}
\label{sec:bif_theta}

\begin{figure}
\centering
\includegraphics[width=.9\textwidth]{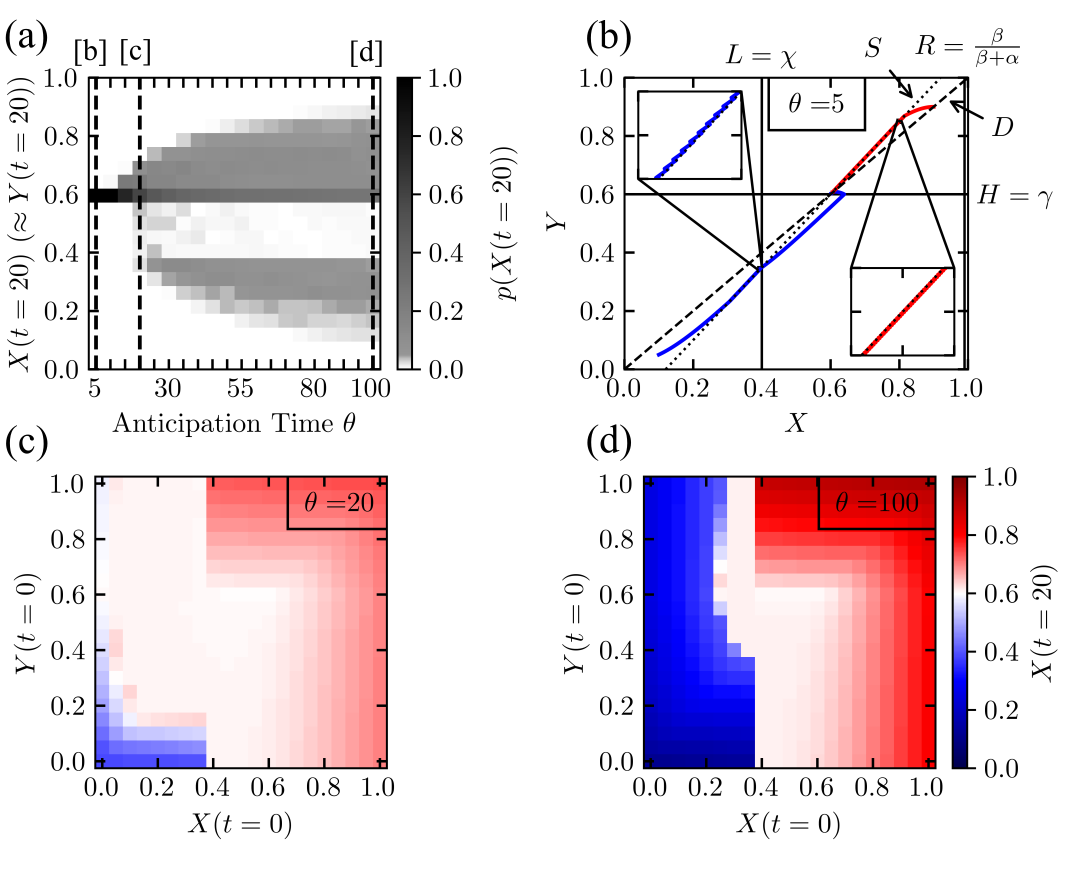}
\caption[Metastability diagram with respect to the anticipation time]{\textbf{(a)}: A metastability diagram with respect to anticipation time $\theta$ is shown for model parameters of $\alpha=0.1$, $\beta=10$, $\tau = 1$, $\gamma=0.6$ and $\chi = 0.4$. Every column corresponds to  a histogram obtained for one $\theta$ representing the occurrences of share of polluters $p(X(t=20))$. The black dashed lines indicate the values of $\theta$ in panels (b--d). \textbf{(b)}: Phase space diagram with two exemplary trajectories (blue and red lines) for $\theta=5$ and other parameters as before. D gives the diagonal where $X=Y$ and $S$  the line where $Y_\text{ant}=\gamma$.  $R$, $L$ and $H$ indicate locations where the trajectories show discontinuous transitions. A more detailed discussion can be found in the Supplement  (see S2 and Fig.~S.1).
\textbf{(c,d)}: Polluting agents $X(t=20)$ as a function of the initial condition plotted as a colour code for  $\theta=20$ \textbf{(c)} and $\theta=100$ \textbf{(d)}. 
}
\label{fig:bif_theta}
\end{figure}

The introduction of anticipation is one of the key features of our model and therefore needs special evaluation. Thus, we show in Fig.\;\ref{fig:bif_theta}\,(a) the long-term share of polluters $p(X(t=20))$ resolved for varying anticipation times and model parameters $\alpha=0.1$, $\beta=10$ and $\tau=1$ as a histogram. For $\alpha>\beta$ the dynamics is dominated by direct environmental impacts and the system converges towards one attractor at the pollution threshold $\gamma$ similar to Fig.\;\ref{fig:case1_2}\;(e--h). The choices of $\alpha=0.1$ and $\beta=10$ ensure that we can observe social dynamics which are driven dominantly by social contagion processes instead of behavioural changes due to direct environmental impacts. Our choice of $\tau$ allows for the dynamic evolution of the environment on similar timescales as the social dynamics, compare Fig.\;\ref{fig:case1_2}\;(e--h).

As our main focus is the qualitative model behaviour we use the mean-field approximation given by Eq.\;(\ref{eq:MF}) which leads to an overall good agreement to the numerical experiments (compare Fig.\;\ref{fig:case1_2}). In Fig.~\ref{fig:bif_theta} the results for the dynamics with different anticipation times $\theta$ are displayed. On the left-hand side of Fig.\;\ref{fig:bif_theta}\,(a) the existance of only one attractor at $X=Y=\gamma=0.6$ is clear.  This reproduces for small $\theta$ the results from the previous section, compare Fig.\;\ref{fig:case1_2}\,(g). 
In Fig.~\ref{fig:bif_theta}\,(b) a phase space diagram is shown for two exemplary trajectories (red and blue lines) in which the trajectories approach the line where the anticipated pollution is equal to $Y_\text{ant} =Y+\frac{\theta}{\tau}(X-Y) =\gamma$ (dotted line $S$ in Fig.~\ref{fig:bif_theta}\,(b)) and then approach the attractor at $X=Y=\gamma$ through a large number of small steps. However, between $\chi$ and $\gamma$ the trajectories can deviate from $S$. In the Supplement (see S2), we show that indeed $X=Y=\gamma$ is a stable fixed point of the deterministic macroscopic approximation whenever $\chi<\gamma<\beta/(\alpha+\beta)$ (see lines $L$, $H$ and $R$ in Fig.~\ref{fig:bif_theta}\,(b) respectively) and $\theta>\tau$.

Despite this unique attractor that determines the long-term fate of the system, two kinds of long metastable behaviour emerge if the anticipation time $\theta$ is large enough. In the snapshot at time $t=20$ shown in Fig.\;\ref{fig:bif_theta}\,(a) as a function of $\theta$, one can see that first an upper ``branch'' appears for $\theta \gtrsim 10$, representing trajectories that at some point visit a region of both high $X$ and $Y$ and then decrease very slowly towards $\gamma$, like the red one shown in Fig.\;\ref{fig:bif_theta}\,(b).
More interestingly, also a lower branch becomes visible 
for $\theta \gtrsim 20$, representing trajectories that at some point visit a region of both low $X$ and $Y$, then increase slowly towards $X=\chi$ and $Y_\text{ant}=\gamma$ before then suddenly moving fast to $X,Y>\gamma$ and eventually converging to the attractor,
like the blue one shown in Fig.\;\ref{fig:bif_theta}\,(b).
The exact dependence of the system's position at $t=20$ on its initial condition is shown in Fig.\;\ref{fig:bif_theta}\,(c,d) for $\theta=20$ and $\theta=100$, red and blue areas representing $X(t=20)>\gamma$ and $X(t=20)<\gamma$, respectively.

The metastability in both branches arises because of two effects. First, close to line $S$, the sign of $\dot X$ alternates fast (as seen in the insets in Fig.\;\ref{fig:case1_2}\;(b)), keeping the system close to that line. Second, the larger $\theta$, the closer line $S$ is to the diagonal $D$, making $|\dot Y|$ become very small when on $S$. Since in the macroscopic approximation, $\dot Y$ has a constant sign throughout such a phase, it slowly converges to either $X=Y=\gamma$ (red) or $X=\chi$, $Y_\text{ant}=\gamma$ (blue). In the latter case, once the system crosses $X=\chi$ (line $L$), it moves fast towards the upper right. In addition to these long-term effects that can be well explained by the macroscopic approximation, the micro-model exhibits even longer metastable phases when $\theta$ becomes large. In that case the lines $S$ and $D$ are close enough so that the finite-size changes of the micro-model can move the system across $D$ while close to $S$. As a result, not only the sign of $\dot X$ but also that of $\dot Y$ will alternate. In the limit of $\theta\to\infty$, $S$ and $D$ coincide and the system may stay forever in any point on $S$ with $X<\chi$ or $\gamma<X<\beta/(\alpha+\beta)$ (see S3 in the Supplement for details).

\section{Discussion \& Conclusion}
\label{sec:disc}

We have proposed a conceptual co-evolutionary model of social contagion, environmental dynamics and associated immediate and anticipated impacts that serve as a mediator between the two subsystems. 
The coupling of the social system to a dynamic environment has led to  a damping of the social dynamics onto one attractor. The influence of anticipation on the dynamics  has opened the possibility to guide the system towards a region with low shares of polluting agents $X$.

In  Section\;\ref{sec:case1_2} we recovered that the pure social dynamics can be multistable if the environment is static. However, introducing a direct coupling to a dynamic environment damped the multistability and one main attractor formed exactly at the pollution threshold $\gamma$. This can be explained by the general tendency of the agents to lower their environmental impact if the pollution is high but also reduce their mitigation cost if the pollution is low. 
If the agents act short-sightedly and adjust their behaviour directly with regard to the current pollution level,
this leads to a fast
convergence towards a trade-off scenario between environmental and ``economic'' benefits. 
Thus, if the pollution threshold were to represent the level of pollution the agents really consider ``optimal'', the short-sighted behaviour would be perfectly rational since it reaches and stabilises the optimum as fast as possible.
In our more ``boundedly rational'' interpretation of the model, however, the pollution threshold does not represent the optimal level of pollution. Instead it is assumed to be much lower, but represents the level at which the problem of pollution is perceived as so urgent by the agent that immediate action seems due. 
For that case, we considered an additional, farsighted driver for behaviour that is however modulated by social dynamics: agents may stop polluting already when they anticipate pollution to cross the pollution threshold at a certain future time and enough of their neighbours are non-polluting already. 
If this additional, neighbour-pertaining threshold $\chi$ is well below the pollution threshold and the anticipation time is high, this has the effect that trajectories starting at low pollution levels become metastable, staying for very long time close to the (lower) neighbour-pertaining threshold. Afterwards, it suddenly increases to a level above the pollution threshold. From here the trajectories eventually converge to the threshold from above (blue line in Fig.\ \ref{fig:bif_theta}b).
The downside of this encouraging result is that now trajectories starting at high pollution levels also become metastable {\em above} the threshold and decrease towards it only very slowly without an intermediate low pollution phase (red lines in Fig.\ \ref{fig:bif_theta}b).
This metastable regime seems to form rather nonlinearly when $\theta$ is increased. Even though Fig.\ \ref{fig:bif_theta}\,(a) reminds of a bifurcation diagram, there is no actual bifurcation: $X=Y=\gamma$ remains the only attractor of the deterministic macroscopic approximation throughout. The stochastic microscopic model, however, can in principle visit and stay in the region around $X=Y=\chi$ frequently due to its finite-size stochastic jumps.  Thus, we argue to find a process that one may denote as social tipping even though  it is not possible to define only one critical anticipation time above which the metastable behaviour predominantly exists. We conclude that the larger the anticipation time, the longer we expect the system to remain in the metastable regime before potentially (at least in the mean-field approximation) converging back to the stable attractor. The desired time that the system should stay within the transient metastable regime may very well vary across different real-world applications. Hence, our results illustrate that social tipping can be interpreted as more than simply a parameter-induced (saddle-node) bifurcation. Instead our model results align with recent definitions of social tipping that specifically account for the existence of qualitatively alternative (meta-)stable states whose existence can not solely be explained by a single control parameter \citep{Milkoreit2018, winkelmann2020social}.

If one wants to decrease the long-term average pollution level of the modelled system by tweaking its parameters or directly influencing its state, one could consider the following strategy.
Whenever the system is at $X>\chi$ (more agents polluting than the neighbour-pertaining threshold), it first has to be moved to $X<\chi$. In the real world, this could be achieved by temporary measures such as taxation or punishment for pollution or subsidies for environmentally friendly behaviours.
Note that a mere reduction of $X$ below $\gamma$ but not below $\chi$, or a reduction of $Y$ instead of $X$ would not suffice since then $X$ and $Y$ would grow again due to the social dynamics.
If the anticipation time $\theta$ can be influenced, then
once the system is at $X<\chi$, $\theta$ could be raised to increase the length of the resulting metastable phase with $X<\chi$. How this could be achieved in reality is less clear, since the anticipation time is strongly linked to how strongly agents care about the future. 
If the neighbour-pertaining threshold $\chi$ can be influenced, a tradeoff emerges between (i) lowering $\chi$ to achieve lower pollution levels during the metastable phase, and (ii) keeping $\chi$ large enough so that $X$ can always be brought back to $X<\chi$ by political measures once a metastable phase has ended.
Note that such political measures seem not essential during a metastable phase but may of course help to prolong it.

Even though the conceptual model is relatively simple, it provides important and surprising insights into the socio-ecological dynamics with anticipation as a key feature. However, it offers room for adjustments, e.g.\ the implementation of a more realistic environmental dynamics which themselves could be subject to tipping such as previously described for ice dynamics \citep{Levermann2016, Klose2020}. 
Studying the effect of logistic growing change  probabilities can alter a systems overall behaviour and lead to novel trajectories \citep{Dodds2005,Traulsen2010} which would be a promising addition to our model as well. 
In general, studying other effects in addition to social activation, such as adaption, which has been shown to  play a crucial role for the stability of co-evolutionary socio-ecological systems \citep{Wiedermann2015,Geier2019,Auer2015,holstein2020}, is a promising step to broaden the application of our model and strengthen its meaningfulness.\\
In summary our work provides a valuable first step in investigating potentially emergent dynamics from interactions between individuals in a dynamic and accessible environment.

\begin{acknowledgement}
The European Regional Development Fund, BMBF and the Land Brandenburg supported this project by providing resources on the high-performance computer system at the Potsdam Institute for Climate Impact Research. M.W. is supported by the Leibniz Association (project DOMINOES).
J.K. acknowledges support from the Russian Ministry of Science and Education ``Digital biodesign and personalised healthcare''.

\end{acknowledgement}

\printbibliography[title={References}]

@article{Bonabeau2002,
%abstract = {Agent-based modeling is a powerful simulation modeling technique that has seen a number of applications in the last few years, including applications to real-world business problems. After the basic principles of agent-based simulation are briefly introduced, its four areas of application are discussed by using real-world applications: flow simulation, organizational simulation, market simulation, and diffusion simulation. For each category, one or several business applications are described and analyzed.ABM, agent-based modelingNASDAQ, National Association of Security Dealers Automated QuotationISP, Internet service provider},
author = {Bonabeau, Eric},
doi = {10.1073/pnas.082080899},
%file = {:home/paul/Masterarbeit/Papers/7280.full.pdf:pdf},
journal = {P.  Natl. Acad Sci. USA},%Proceedings of the National Academy of Sciences}

@article{Hansen1981,
%abstract = {The global temperature rose by 0.2°C between the middle 1960's and 1980, yielding a warming of 0.4°C in the past century. This temperature increase is consistent with the calculated greenhouse effect due to measured increases of atmospheric carbon dioxide. Variations of volcanic aerosols and possibly solar luminosity appear to be primary causes of observed fluctuations about the mean trend of increasing temperature. It is shown that the anthropogenic carbon dioxide warming should emerge from the noise level of natural climate variability by the end of the century, and there is a high probability of warming in the 1980's. Potential effects on climate in the 21st century include the creation of drought-prone regions In North America and central Asia as part of a shifting of climatic zones, erosion of the West Antarctic ice sheet with a consequent worldwide rise in sea level, and opening of the fabled Northwest Passage. Copyright {\textcopyright} 1981 AAAS.},
author = {Hansen, J. and Johnson, D. and Lacis, A. and Lebedeff, S. and Lee, P. and Rind, D. and Russell, G.},
doi = {10.1126/science.213.4511.957},
%file = {:home/paul/Masterarbeit/Papers/1981{\_}Hansen{\_}ha04600x.pdf:pdf},
%issn = {00368075},
journal = {Science},
number = {4511},
pages = {957--966},
title = {{Climate impact of increasing atmospheric carbon dioxide}},
volume = {213},
year = {1981}
}

@article{Stott2000,
%abstract = {A comparison of observations with simulations of a coupled ocean-atmosphere general circulation model shows that both natural and anthropogenic factors have contributed significantly to 20th century temperature changes. The model successfully simulates global mean and large-scale land temperature variations, indicating that the climate response on these scales is strongly influenced by external factors. More than 80{\%} of observed multidecadal-scale global mean temperature variations and more than 60{\%} of 10- to 50-year land temperature variations are due to changes in external forcings. Anthropogenic global warming under a standard emissions scenario is predicted to continue at a rate similar to that observed in recent decades.},
author = {Stott, P. A. and Tett, S. F.B. and Jones, G. S. and Allen, M. R. and Mitchell, J. F.B. and Jenkins, G. J.},
doi = {10.1126/science.290.5499.2133},
%file = {:home/paul/Masterarbeit/Papers/2133.full.pdf:pdf},
%issn = {00368075},
journal = {Science},
number = {5499},
pages = {2133--2137},
title = {{External control of 20th century temperature by natural and anthropogenic forcings}},
volume = {290},
year = {2000}
}

@article{Kellie-Smith2011,
%abstract = {Global CO2 emissions are understood to be the largest contributor to anthropogenic climate change, and have, to date, been highly correlated with economic output. However, there is likely to be a negative feedback between climate change and human wealth: economic growth is typically associated with an increase in CO2 emissions and globalwarming, but the resulting climate change may lead to damages that suppress economic growth. This climate-economy feedback is assumed to be weak in standard climate change assessments. When the feedback is incorporated in a transparently simple model it reveals possible emergent behaviour in the coupled climate-economy system. Formulae are derived for the critical rates of growth of global CO2 emissions that cause damped or long-term boom-bust oscillations in human wealth, thereby preventing a soft landing of the climate-economy system. On the basis of this model, historical rates of economic growth and decarbonization appear to put the climate-economy system in a potentially damaging oscillatory regime. {\textcopyright} 2011 The Royal Society.},
author = {Kellie-Smith, Owen and Cox, Peter M.},
doi = {10.1098/rsta.2010.0305},
%file = {:home/paul/Masterarbeit/Papers/Kellie-Smith, Cox - 2011 - Emergent dynamics of the climate economy system in the Anthropocene.pdf:pdf},
%issn = {1364503X},
journal = {Philos. T. Roy. Soc. A},%ophical Transactions of the Royal Society A: Mathematical, Physical and Engineering Sciences}

@unpublished{winkelmann2020social,
      title={Social tipping processes for sustainability: An analytical framework}, 
      author={Ricarda Winkelmann and Jonathan F. Donges and E. Keith Smith and Manjana Milkoreit and Christina Eder and Jobst Heitzig and Alexia Katsanidou and Marc Wiedermann and Nico Wunderling and Timothy M. Lenton},
      year={2020},
      eprint={arXiv:2010.04488},
      archivePrefix={arXiv},
      primaryClass={physics.soc-ph}
}

@article{GranThres,
author = {Granovetter, Mark},
doi = {10.1007/978-3-658-21742-6_54},
%file = {:home/paul/Masterarbeit/Papers/granthreshold.pdf:pdf},
journal = {Am. J. Sociol.},
number = {6},
pages = {1420--1443},
title = {{Threshold models of collective behavior}},
volume = {83},
year = {1978}
}

@article{Wiedermann2020,
%abstract = {Social tipping, where minorities trigger larger populations to engage in collective action, has been suggested as one key aspect in addressing contemporary global challenges. Here, we refine Granovetter's widely acknowledged theoretical threshold model of collective behavior as a numerical modelling tool for understanding social tipping processes and resolve issues that so far have hindered such applications. Based on real-world observations and social movement theory, we group the population into certain or potential actors, such that -- in contrast to its original formulation -- the model predicts non-trivial final shares of acting individuals. Then, we use a network cascade model to explain and analytically derive that previously hypothesized broad threshold distributions emerge if individuals become active via social interaction. Thus, through intuitive parameters and low dimensionality our refined model is adaptable to explain the likelihood of engaging in collective behavior where social tipping like processes emerge as saddle-node bifurcations and hysteresis.},
%archivePrefix = {arXiv},
%arxivId = {1911.04126},
author = {Wiedermann, Marc and Smith, E. Keith and Heitzig, Jobst and Donges, Jonathan F.},
journal = {Sci. Rep.},
%eprint = {1911.04126},
%file = {:home/paul/Masterarbeit/Papers/1911.04126.pdf:pdf},
pages = {1--23},
title = {{A network-based microfoundation of Granovetter's threshold model for social tipping}},
%url = {http://arxiv.org/abs/1911.04126},
year = {2020},
volume = {10},
number={11202}
}

@article{Muller-hansen2017,
author = {M{\"{u}}ller-Hansen, Finn and Schl{\"{u}}ter, Maja and M{\"{a}}s, Michael and Donges, Jonathan F and Kolb, Jakob J and Thonicke, Kirsten and Heitzig, Jobst},
%file = {:home/paul/Masterarbeit/Papers/esd-8-977-2017.pdf:pdf},
journal = {Earth Syst. Dynam},
pages = {977--1007},
title = {{Towards representing human behavior and decision making in Earth system models – an overview of techniques and approaches}},
volume = {8},
year = {2017},
number = {1}
}

@unpublished{Winkelmann2019,
%abstract = {Vital parts of the climate system such as the West Antarctic Ice Sheet are at risk even within the aspired aims of the Paris Agreement to limit global temperature rise to 1.5-2 • C. Implementing effective climate policies critically depends on cascading effects from (anticipated) climate impacts to emission reductions via social tipping.},
%archivePrefix = {arXiv},
%arxivId = {arXiv:1911.10063v2},
author = {Wiedermann, Marc and Winkelmann, Ricarda and Donges, Jonathan F and Eder, Christina and Heitzig, Jobst and Katsanidou, Alexia and Smith, E Keith and Donges, Jonathan F and Eder, Christina and Heitzig, Jobst and Katsanidou, Alexia and {Keith Smith}, E},
eprint = {arXiv:1911.10063v2},
%file = {:home/paul/Masterarbeit/Papers/1911.10063.pdf:pdf;:home/paul/Masterarbeit/Papers/1911.10063.pdf:pdf},
%keywords = {,author for correspondence,climate,climate action,climate concern,climate impacts,concern,social tipping},
pages = {1--5},
title = {{Domino Effects in the Earth System – The potential role of wanted tipping points}},
year = {2019}
}

@article{Milkoreit2018,
%abstract = {The term tipping point has experienced explosive popularity across multiple disciplines over the last decade. Research on social-ecological systems (SES) has contributed to the growth and diversity of the term's use. The diverse uses of the term obscure potential differences between tipping behavior in natural and social systems, and issues of causality across natural and social system components in SES. This paper aims to create the foundation for a discussion within the SES research community about the appropriate use of the term tipping point, especially the relatively novel term 'social tipping point.' We review existing literature on tipping points and similar concepts (e.g. regime shifts, critical transitions) across all spheres of science published between 1960 and 2016 with a special focus on a recent and still small body of work on social tipping points. We combine quantitative and qualitative analyses in a bibliometric approach, rooted in an expert elicitation process. We find that the term tipping point became popular after the year 2000 - long after the terms regime shift and critical transition - across all spheres of science. We identify 23 distinct features of tipping point definitions and their prevalence across disciplines, but find no clear taxonomy of discipline-specific definitions. Building on the most frequently used features, we propose definitions for tipping points in general and social tipping points in SES in particular.},
author = {Milkoreit, Manjana and Hodbod, Jennifer and Baggio, Jacopo and Benessaiah, Karina and Calder{\'{o}}n-Contreras, Rafael and Donges, Jonathan F. and Mathias, Jean Denis and Rocha, Juan Carlos and Schoon, Michael and Werners, Saskia E.},
doi = {10.1088/1748-9326/aaaa75},
%file = {:home/paul/Masterarbeit/Papers/Milkoreit{\_}2018{\_}Environ.{\_}Res.{\_}Lett.{\_}13{\_}033005.pdf:pdf},
%issn = {17489326},
journal = {Environ. Res. Lett.},
%keywords = {non-linear change,social tipping points,social-ecological systems,tipping points},
number = {3},
title = {{Defining tipping points for social-ecological systems scholarship - An interdisciplinary literature review}},
volume = {13},
year = {2018}
}

@article{Xie2011,
author = {Xie, J and Sreenivasan, S and Korniss, G and Zhang, W and Lim, C and Szymanski, B K},
doi = {10.1103/PhysRevE.84.011130},
%file = {:home/paul/Masterarbeit/Papers/PhysRevE.84.011130.pdf:pdf},
journal = {Phys. Rev. E},
pages = {011130},
title = {{Social consensus through the influence of committed minorities}},
volume = {84},
year = {2011},
number = {1}
}

@article{Macy2002,
author = {Macy, Michael W and Willer, Robert},
doi = {10.1146/annurev.soc.28.110601.141117},
%file = {:home/paul/Masterarbeit/Papers/3069238.pdf:pdf},
journal = {Annu. Rev. Sociol.},
%keywords = {ables,abstract sociologists often model,among vari-,approach that models social,automata,cellular,complexity,emergence,genetic algorithm,life as interactions among,self-organization,simulation,social processes as interactions,we review an alternative},
pages = {143--166},
title = {{From factors to actors : Computational sociology and agent-based modeling}},
volume = {28},
year = {2002},
number = {1}
}

@article{Watts2002,
author = {Watts, Duncan J},
doi = {10.1073/082090499},
%file = {:home/paul/Masterarbeit/Papers/cascades.pdf:pdf},
%issn = {0027-8424},
journal = {P.  Natl. Acad Sci. USA},%Proceedings of the National Academy of Sciences}

@article{Donges2018a,
%abstract = {Abstract. In the Anthropocene, social processes have become critical to understanding planetary-scale Earth system dynamics. The conceptual foundations of Earth system modelling have externalised social processes in ways that now hinder progress in understanding Earth resilience and informing governance of global environmental change. New approaches to global modelling are needed to address these challenges, but the current modelling landscape is highly diverse and heterogeneous, ranging from purely biophysical Earth System Models, to hybrid macro-economic Integrated Assessments Models, to a plethora of models of socio-cultural dynamics. World-Earth models, currently not yet available, will need to integrate all these elements, so future World-Earth modellers require a structured approach to identify, classify, select, and combine model components. Here, we develop taxonomies for ordering the multitude of societal and biophysical subsystems and their interactions. We suggest three taxa for modelled subsystems: (i) biophysical, where dynamics is usually represented by natural laws of physics, chemistry or ecology (i.e., the usual components of Earth system models), (ii) socio-cultural, dominated by processes of human behaviour, decision making and collective social dynamics (e.g., politics, institutions, social networks), and (iii) socio-metabolic, dealing with the material interactions of social and biophysical subsystems (e.g., human bodies, natural resource and agriculture). We show how higher-order taxonomies for interactions between two or more subsystems can be derived, highlighting the kinds of social-ecological feedback loops where new modelling efforts need to be directed. As an example, we apply the taxonomy to a stylised World-Earth system model of socially transmitted discount rates in a greenhouse gas emissions game to illustrate the effects of social-ecological feedback loops that are usually not considered in current modelling efforts. The proposed taxonomy can contribute to guiding the design and operational development of more comprehensive World-Earth models for understanding Earth resilience and charting sustainability transitions within planetary boundaries and other future trajectories in the Anthropocene.},
%annote = {Goldgrube an Citations},
author = {Donges, Jonathan F. and Lucht, Wolfgang and Heitzig, Jobst and Barfuss, Wolfram and Cornell, Sarah E. and Lade, Steven J. and Schl{\"{u}}ter, Maja},
doi = {10.5194/esd-2018-27},
%file = {:home/paul/Masterarbeit/Papers/esd-2018-27.pdf:pdf},
%issn = {2190-4995},
journal = {Earth Syst. Dynam. Discuss.},
title = {{Taxonomies for structuring models for World-Earth system analysis of the Anthropocene: subsystems, their interactions and social-ecological feedback loops}},
volume = {2018},
%number = {in review},
year = {2018}
}

@article{Donges2018,
%abstract = {Abstract. Possible future trajectories of the Earth system in the Anthropocene are determined by the increasing entanglement of processes operating in the physical, chemical and biological systems of the planet, as well as in human societies, their cultures and economies. Here, we introduce the copan:CORE open source software library that provides a framework for developing, composing and running World-Earth models, i.e., models of social-ecological co-evolution up to planetary scales. It is an object-oriented software package written in Python designed for different user roles. It allows model end users to run parallel simulations with already available and tested models. Furthermore, model composers are enabled to easily implement new models by plugging together a broad range of model components, such as opinion formation on social networks, generic carbon cycle dynamics, or simple vegetation growth. For the sake of a modular structure, each provided component specifies a meaningful yet minimal collection of closely related processes. These processes can be formulated in terms of various process types, such as ordinary differential equations, explicit or implicit functions, as well as steps or events of deterministic or stochastic fashion. In addition to the already included variety of different components in copan:CORE, model developers can extend the framework with additional components that are based on elementary entity types, i.e., grid cells, individuals and social systems, or the fundamental process taxa environment, social metabolism, and culture. To showcase possible usage we present an exemplary World-Earth model that combines a variety of model components and interactions thereof. As the framework allows a simple activation and deactivation of certain components and related processes, users can test for their specific effects on modeling results and evaluate model robustness in a controlled way. Hence, copan:CORE allows developing process-based models of global change and sustainable development in planetary social-ecological systems and thus fosters a better understanding of crucial mechanisms governing the co-evolutionary dynamics between societies and the natural environment. Due to its modular structure, the framework enhances the development and application of stylized models in Earth system science but also climatology, economics, ecology, or sociology, and allows combining them for interdisciplinary studies at the interface between different areas of expertise.},
author = {{J.F. Donges and J. Heitzig} and Barfuss, Wolfram and Kassel, Johannes A. and Kittel, Tim and Kolb, Jakob J. and Kolster, Till and M{\"{u}}ller-Hansen, Finn and Otto, Ilona M. and Wiedermann, Marc and Zimmerer, Kilian B. and Lucht, Wolfgang},
doi = {10.5194/esd-2017-126},
%file = {:home/paul/Masterarbeit/Papers/esd-2017-126.pdf:pdf},
journal = {Earth Syst. Dynam},
volume = {11},
%number = {1},
pages = {1--27},
title = {{Earth system modelling with complex dynamic human societies: the copan:CORE World-Earth modeling framework}},
year = {2020}
}

@article{Lenton2008,
author = {Lenton, Timothy M. and Held, Hermann and Kriegler, Elmar and Hall, Jim W. and Lucht, Wolfgang and Rahmstorf, Stefan and Schellnhuber, Hans Joachim},
doi = {10.1073/pnas.0911106106},
%file = {:home/paul/Masterarbeit/Papers/1786.full.pdf:pdf},
%issn = {10916490},
journal = {P.  Natl. Acad Sci. USA},%Proceedings of the National Academy of Sciences}

@article{Centola2018,
%abstract = {Theoretical models of critical mass have shown how minority groups can initiate social change dynamics in the emergence of new social conventions. Here, we study an artificial system of social conventions in which human subjects interact to establish a new coordination equilibrium. The findings provide direct empirical demonstration of the existence of a tipping point in the dynamics of changing social conventions. When minority groups reached the critical mass-that is, the critical group size for initiating social change-they were consistently able to overturn the established behavior. The size of the required critical mass is expected to vary based on theoretically identifiable features of a social setting. Our results show that the theoretically predicted dynamics of critical mass do in fact emerge as expected within an empirical system of social coordination.},
author = {Centola, Damon and Becker, Joshua and Brackbill, Devon and Baronchelli, Andrea},
doi = {10.1126/science.aas8827},
%file = {:home/paul/Masterarbeit/Papers/1116.full.pdf:pdf},
%issn = {10959203},
journal = {Science},
number = {6393},
pages = {1116--1119},
title = {{Experimental evidence for tipping points in social convention}},
volume = {360},
year = {2018}
}

@article{Schellnhuber2016,
author = {Schellnhuber, Hans Joachim and Rahmstorf, Stefan and Winkelmann, Ricarda},
doi = {10.1038/nclimate3013},
%file = {:home/paul/Masterarbeit/Papers/nclimate3013.pdf:pdf},
%issn = {17586798},
journal = {Nat. Clim. Change},
number = {7},
pages = {649--653},
title = {{Why the right climate target was agreed in Paris}},
volume = {6},
year = {2016}
}

@article{Kriegler2009,
abstract = {Major restructuring of the Atlantic meridional overturning circulation, the Greenland and West Antarctic ice sheets, the Amazon rainforest and ENSO, are a source of concern for climate policy. We have elicited subjective probability intervals for the occurrence of such major changes under global warming from 43 scientists. Although the expert estimates highlight large uncertainty, they allocate significant probability to some of the events listed above. We deduce conservative lower bounds for the probability of triggering at least 1 of those events of 0.16 for medium (2-4 °C), and 0.56 for high global mean temperature change (above 4 °C) relative to year 2000 levels.},
author = {Kriegler, Elmar and Hall, Jim W. and Held, Hermann and Dawson, Richard and Schellnhuber, Hans Joachim},
doi = {10.1073/pnas.0809117106},
%file = {:home/paul/Masterarbeit/Papers/5041.full.pdf:pdf},
%issn = {00278424},
journal = {P.  Natl. Acad Sci. USA},
%keywords = {Climate change,xpert elicitation},
number = {13},
pages = {5041--5046},
title = {{Imprecise probability assessment of tipping points in the climate system}},
volume = {106},
year = {2009}
}

@article{Sabine2004,
%annote = {-basically the ocean is the only net carbon sink},
author = {Sabine, Chrisopher L. and Feely, Richard A. and Gruber, Nicolas and Key, Robert M. and Lee, Kitack and Bullister, John L. and Wanninkhof, Rik and Wong, C. S. and Wallace, Douglas W. R. and Tilbrook, Bronte and Millero, Frank J. and Peng, Tsung-Hung and Kozyr, Alexander and Ono, Tuseno and Rios, Aida F.},
doi = {doi:10.1126/science.1097403},
%file = {:home/paul/Masterarbeit/Papers/367.full.pdf:pdf},
journal = {Science},
number = {5682},
pages = {367--371},
title = {{The cceanic sink for anthropogenic CO2}},
volume = {305},
year = {2004}
}

@article{Cox2000,
%annote = {- At 2000 around half of carbon (fossil burned) is sotred away by ocean.},
author = {Cox, Peter M. and Betts, Richard A and Jones, Chris D and Spall, Steven A and Totterdell, Ian J.},
%file = {:home/paul/Masterarbeit/Papers/35041539.pdf:pdf},
journal = {Nature},
pages = {184--187},
title = {{Acceleration of global warming due to carbon-cycle feedbacks in a coupled climate model}},
volume = {408},
year = {2000},
number = {6809}
}

@article{Moore1994,
abstract = {We explore the effects of a changing terrestrial biosphere on the atmospheric residence time of CO2 using three simple ocean carbon cycle models and a model of global terrestrial carbon cycling. We find differences in model behavior associated with the assumption of an active terrestrial biosphere (forest regrowth) and significant differences if we assume a donor-dependent flux from the atmosphere to the terrestrial component (e.g., a hypothetical terrestrial fertilization flux). To avoid numerical difficulties associated with treating the atmospheric CO2 decay (relaxation) curve as being well approximated by a weighted sum of exponential functions, we define the single half-life as the time it takes for a model atmosphere to relax from its present-day value half way to its equilibrium pCO2 value. This scenario-based approach also avoids the use of unit pulse (Dirac Delta) functions which can prove troublesome or unrealistic in the context of a terrestrial fertilization assumption. We also discuss some of the numerical problems associated with a conventional lifetime calculation which is based on an exponential model. We connect our analysis of the residence time of CO2 and the concept of single half-life to the residence time calculations which are based on using weighted sums of exponentials. We note that the single half-life concept focuses upon the early decline of CO2 under a cutoff/decay scenario. If one assumes a terrestrial biosphere with a fertilization flux, then our best estimate is that the single half-life for excess CO2 lies within the range of 19 to 49 years, with a reasonable average being 31 years. If we assume only regrowth, then the average value for the single half-life for excess CO2 increases to 72 years, and if we remove the terrestrial component completely, then it increases further to 92 years.},
annote = {CO{\_}2 atmospheric surplus half life time = 19-49 years
e{\_}time = 27-70 years},
author = {Moore, Berrien and Braswell, B. H.},
doi = {10.1029/93GB03392},
%file = {:home/paul/Masterarbeit/Papers/53b7067d53eb35ddb6085e57e8a6461e24b8.pdf:pdf},
%issn = {19449224},
journal = {Global Biogeochem. Cy.},
number = {1},
pages = {23--38},
title = {{The lifetime of excess atmospheric carbon dioxide}},
volume = {8},
year = {1994}
}

@article{Otto2020,
%abstract = {Safely achieving the goals of the Paris Climate Agreement requires a worldwide transformation to carbon-neutral societies within the next 30 y. Accelerated technological progress and policy implementations are required to deliver emissions reductions at rates sufficiently fast to avoid crossing dangerous tipping points in the Earth's climate system. Here, we discuss and evaluate the potential of social tipping interventions (STIs) that can activate contagious processes of rapidly spreading technologies, behaviors, social norms, and structural reorganization within their functional domains that we refer to as social tipping elements (STEs). STEs are subdomains of the planetary socioeconomic system where the required disruptive change may take place and lead to a sufficiently fast reduction in anthropogenic greenhouse gas emissions. The results are based on online expert elicitation, a subsequent expert workshop, and a literature review. The STIs that could trigger the tipping of STE subsystems include 1) removing fossil-fuel subsidies and incentivizing decentralized energy generation (STE1, energy production and storage systems), 2) building carbon-neutral cities (STE2, human settlements), 3) divesting from assets linked to fossil fuels (STE3, financial markets), 4) revealing the moral implications of fossil fuels (STE4, norms and value systems), 5) strengthening climate education and engagement (STE5, education system), and 6) disclosing information on greenhouse gas emissions (STE6, information feedbacks). Our research reveals important areas of focus for larger-scale empirical and modeling efforts to better understand the potentials of harnessing social tipping dynamics for climate change mitigation.},
author = {Otto, Ilona M. and Donges, Jonathan F. and Cremades, Roger and Bhowmik, Avit and Hewitt, Richard J. and Lucht, Wolfgang and Rockstr{\"{o}}m, Johan and Allerberger, Franziska and McCaffrey, Mark and Doe, Sylvanus S.P. and Lenferna, Alex and Mor{\'{a}}n, Nerea and van Vuuren, Detlef P. and Schellnhuber, Hans Joachim},
doi = {10.1073/pnas.1900577117},
%file = {:home/paul/Masterarbeit/Papers/2354.full.pdf:pdf},
%issn = {10916490},
journal = {P.  Natl. Acad Sci. USA},%Proceedings of the National Academy of Sciences}

@article{Donges2017,
%abstract = {International commitment to the appropriately ambitious Paris climate agreement and the United Nations Sustainable Development Goals in 2015 has pulled into the limelight the urgent need for major scientific progress in understanding and modelling the Anthropocene, the tightly intertwined social-environmental planetary system that humanity now inhabits. The Anthropocene qualitatively differs from previous eras in Earth's history in three key characteristics: (1) There is planetary-scale human agency. (2) There are social and economic networks of teleconnections spanning the globe. (3) It is dominated by planetary-scale social-ecological feedbacks. Bolting together old concepts and methodologies cannot be an adequate approach to describing this new geological era. Instead, we need a new paradigm in Earth System science that is founded equally on a deep understanding of the physical and biological Earth System – and of the economic, social and cultural forces that are now an intrinsic part of it. It is time to close the loop and bring socially mediated dynamics explicitly into theory, analysis and models that let us study the whole Earth System.},
author = {Donges, Jonathan F. and Winkelmann, Ricarda and Lucht, Wolfgang and Cornell, Sarah E. and Dyke, James G. and Rockstr{\"{o}}m, Johan and Heitzig, Jobst and Schellnhuber, Hans Joachim},
doi = {10.1177/2053019617725537},
%file = {:home/paul/Masterarbeit/Papers/2053019617725537.pdf:pdf},
%issn = {2053020X},
journal = {Anthropocene Review},
%keywords = {Earth System analysis,Earth System modelling,coevolutionary dynamics,complex adaptive networks,human agency,planetary boundaries,safe and just space for humanity,sustainable development goals},
number = {2},
pages = {151--157},
title = {{Closing the loop: Reconnecting human dynamics to Earth System science}},
volume = {4},
year = {2017}
}

@article{Page2015,
author = {Page, Scott E},
doi = {10.1146/annurev-soc-073014-112230},
%file = {:home/paul/Masterarbeit/Papers/annurev-soc-073014-112230.pdf:pdf},
journal = {Annu. Rev. Sociol.},
%keywords = {complexity,diversity,methodology,modeling,networks},
pages = {21--41},
title = {{What sociologists should know about complexity}},
volume = {41},
year = {2015},
number = {1}
}

@article{Steffen2011,
%abstract = {The human imprint on the global environment has now become so large and active that it rivals some of the great forces of Nature in its impact on the functioning of the Earth system. Although global-scale human influence on the environment has been recognized since the 1800s, the term Anthropocene, introduced about a decade ago, has only recently become widely, but informally, used in the global change research community. However, the term has yet to be accepted formally as a new geological epoch or era in Earth history. In this paper, we put forward the case for formally recognizing the Anthropocene as a new epoch in Earth history, arguing that the advent of the Industrial Revolution around 1800 provides a logical start date for the new epoch. We then explore recent trends in the evolution of the Anthropocene as humanity proceeds into the twenty-first century, focusing on the profound changes to our relationship with the rest of the living world and on early attempts and proposals for managing our relationship with the large geophysical cycles that drive the Earth's climate system. {\textcopyright} 2011 The Royal Society.},
author = {Steffen, Will and Grinevald, Jacques and Crutzen, Paul and Mcneill, John},
doi = {10.1098/rsta.2010.0327},
%file = {:home/paul/Masterarbeit/Papers/rsta.2010.0327.pdf:pdf},
%issn = {1364503X},
journal = {Philos. T. Roy. Soc. A},%ophical Transactions of the Royal Society A: Mathematical, Physical and Engineering Sciences}

@article{Crutzen2002,
author = {Crutzen, Paul J.},
doi = {10.1038/415023a},
%file = {:home/paul/Masterarbeit/Papers/415023a.pdf:pdf},
%issn = {00280836},
journal = {Nature},
number = {6867},
pages = {23},
title = {{Geology of mankind}},
volume = {415},
year = {2002}
}

@article{Epstein1999,
author = {Epstein, Joshua M},
%file = {:home/paul/Masterarbeit/Papers/(SICI)1099-0526(199905{\_}06)4{\_}5{\_}41{\_}{\_}AID-CPLX9{\_}3.0.CO$\backslash$;2-F.pdf:pdf},
journal = {Complexity},
%keywords = {1,agent-based models,artificial societies,generative social science,philosophy of social science,s question,the generativist},
number = {5},
pages = {41--60},
title = {{Agent-based computational models and generative social science}},
volume = {4},
year = {1999}
}

@article{Bury2019,
%abstract = {Geophysical models of climate change are becoming increasingly sophisticated, yet less effort is devoted to modelling the human systems causing climate change and how the two systems are coupled. Here, we develop a simple socio-climate model by coupling an Earth system model to a social dynamics model. We treat social processes endogenously— emerging from rules governing how individuals learn socially and how social norms develop —as well as being influenced by climate change and mitigation costs. Our goal is to gain qualitative insights into scenarios of potential socio-climate dynamics and to illustrate how such models can generate new research questions. We find that the social learning rate is strongly influential, to the point that variation of its value within empirically plausible ranges changes the peak global temperature anomaly by more than 1˚C. Conversely, social norms reinforce majority behaviour and therefore may not provide help when we most need it because they suppress the early spread of mitigative behaviour. Finally, exploring the model's parameter space for mitigation cost and social learning suggests optimal intervention pathways for climate change mitigation. We find that prioritising an increase in social learning as a first step, followed by a reduction in mitigation costs provides the most efficient route to a reduced peak temperature anomaly. We conclude that socio-climate models should be included in the ensemble of models used to project climate change.},
%annote = {t{\_}p = 0 and t{\_}f/t{\_}p = $\backslash$theta in my model (eq (8))},
author = {Bury, Thomas M. and Bauch, Chris T. and Anand, Madhur},
doi = {10.1371/journal.pcbi.1007000},
%file = {:home/paul/Masterarbeit/Papers/pcbi.1007000.pdf:pdf},
%isbn = {1111111111},
%issn = {15537358},
journal = {PLOS Comput. Biol.},
number = {6},
pages = {1--16},
title = {{Charting pathways to climate change mitigation in a coupled socio-climate model}},
volume = {15},
year = {2019}
}

@article{Palmer2014,
author = {Palmer, Paul I. and Smith, Matthew J.},
doi = {10.1038/512365a},
%file = {:home/paul/Masterarbeit/Papers/512365a.pdf:pdf},
%issn = {14764687},
journal = {Nature},
number = {7515},
pages = {365--366},
title = {{Earth systems: Model human adaptation to climate change}},
volume = {512},
year = {2014}
}

@article{Steffen2018,
abstract = {We explore the risk that self-reinforcing feedbacks could push the Earth System toward a planetary threshold that, if crossed, could prevent stabilization of the climate at intermediate temperature rises and cause continued warming on a “Hothouse Earth” pathway even as human emissions are reduced. Crossing the threshold would lead to a much higher global average temperature than any interglacial in the past 1.2 million years and to sea levels significantly higher than at any time in the Holocene. We examine the evidence that such a threshold might exist and where it might be. If the threshold is crossed, the resulting trajectory would likely cause serious disruptions to ecosystems, society, and economies. Collective human action is required to steer the Earth System away from a potential threshold and stabilize it in a habitable interglacial-like state. Such action entails stewardship of the entire Earth System—biosphere, climate, and societies—and could include decarbonization of the global economy, enhancement of biosphere carbon sinks, behavioral changes, technological innovations, new governance arrangements, and transformed social values.},
author = {Steffen, Will and Rockstr{\"{o}}m, Johan and Richardson, Katherine and Lenton, Timothy M. and Folke, Carl and Liverman, Diana and Summerhayes, Colin P. and Barnosky, Anthony D. and Cornell, Sarah E. and Crucifix, Michel and Donges, Jonathan F. and Fetzer, Ingo and Lade, Steven J. and Scheffer, Marten and Winkelmann, Ricarda and Schellnhuber, Hans Joachim},
doi = {10.1073/pnas.1810141115},
%file = {:home/paul/Masterarbeit/Papers/8252.full.pdf:pdf},
%issn = {10916490},
journal = {P.  Natl. Acad Sci. USA},%Proceedings of the National Academy of Sciences of the United States of America}

@article{Auer2015,
abstract = {Complex networks describe the structure of many socio-economic systems. However, in studies of decision-making processes the evolution of the underlying social relations are disregarded. In this report, we aim to understand the formation of self-organizing domains of cooperation ( € coalitions €) on an acquaintance network. We include both the network € s influence on the formation of coalitions and vice versa how the network adapts to the current coalition structure, thus forming a social feedback loop. We increase complexity from simple opinion adaptation processes studied in earlier research to more complex decision-making determined by costs and benefits, and from bilateral to multilateral cooperation. We show how phase transitions emerge from such coevolutionary dynamics, which can be interpreted as processes of great transformations. If the network adaptation rate is high, the social dynamics prevent the formation of a grand coalition and therefore full cooperation. We find some empirical support for our main results: Our model develops a bimodal coalition size distribution over time similar to those found in social structures. Our detection and distinguishing of phase transitions may be exemplary for other models of socio-economic systems with low agent numbers and therefore strong finite-size effects.},
author = {Auer, S. and Heitzig, J. and Kornek, U. and Sch{\"{o}}ll, E. and Kurths, J.},
doi = {10.1038/srep13386},
%file = {:home/paul/Masterarbeit/Papers/srep13386.pdf:pdf},
%issn = {20452322},
journal = {Sci. Rep.},
pages = {1--8},
title = {{The Dynamics of Coalition Formation on Complex Networks}},
volume = {5},
year = {2015},
number = {13386}
}

@article{Wiedermann2015,
%abstract = {In many real-world complex systems, the time evolution of the network's structure and the dynamic state of its nodes are closely entangled. Here we study opinion formation and imitation on an adaptive complex network which is dependent on the individual dynamic state of each node and vice versa to model the coevolution of renewable resources with the dynamics of harvesting agents on a social network. The adaptive voter model is coupled to a set of identical logistic growth models and we mainly find that, in such systems, the rate of interactions between nodes as well as the adaptive rewiring probability are crucial parameters for controlling the sustainability of the system's equilibrium state. We derive a macroscopic description of the system in terms of ordinary differential equations which provides a general framework to model and quantify the influence of single node dynamics on the macroscopic state of the network. The thus obtained framework is applicable to many fields of study, such as epidemic spreading, opinion formation, or socioecological modeling.},
%archivePrefix = {arXiv},
%arxivId = {1503.05914},
author = {Wiedermann, Marc and Donges, Jonathan F. and Heitzig, Jobst and Lucht, Wolfgang and Kurths, J{\"{u}}rgen},
doi = {10.1103/PhysRevE.91.052801},
%eprint = {1503.05914},
%file = {:home/paul/Masterarbeit/Papers/1503.05914.pdf:pdf},
%issn = {15502376},
journal = {Phys. Rev. E},% - Statistical, Nonlinear, and Soft Matter Physics}

@article{Barfuss2017,
%abstract = {Human societies depend on the resources ecosystems provide. Particularly since the last century, human activities have transformed the relationship between nature and society at a global scale. We study this coevolutionary relationship by utilizing a stylized model of private resource use and social learning on an adaptive network. The latter process is based on two social key dynamics beyond economic paradigms: Boundedly rational imitation of resource use strategies and homophily in the formation of social network ties. The private and logistically growing resources are harvested with either a sustainable (small) or non-sustainable (large) effort. We show that these social processes can have a profound influence on the environmental state, such as determining whether the private renewable resources collapse from overuse or not. Additionally, we demonstrate that heterogeneously distributed regional resource capacities shift the critical social parameters where this resource extraction system collapses. We make these points to argue that, in more advanced coevolutionary models of the planetary social-ecological system, such socio-cultural phenomena as well as regional resource heterogeneities should receive attention in addition to the processes represented in established Earth system and integrated assessment models.},
author = {Barfuss, Wolfram and Donges, F. Jonathan and Wiedermann, Marc and Lucht, Wolfgang},
doi = {10.5194/esd-8-255-2017},
%file = {:home/paul/Masterarbeit/Papers/esd-8-255-2017.pdf:pdf},
%issn = {21904987},
journal = {Earth Syst. Dynam},
number = {2},
pages = {255--264},
title = {{Sustainable use of renewable resources in a stylized social-ecological network model under heterogeneous resource distribution}},
volume = {8},
year = {2017}
}

@article{Nitzbon2017,
%abstract = {The Anthropocene is characterized by close interdependencies between the natural Earth system and the global human society, posing novel challenges to model development. Here we present a conceptual model describing the long-term co-evolution of natural and socio-economic subsystems of Earth. While the climate is represented via a global carbon cycle, we use economic concepts to model socio-metabolic flows of biomass and fossil fuels between nature and society. A well-being-dependent parametrization of fertility and mortality governs human population dynamics. Our analysis focuses on assessing possible asymptotic states of the Earth system for a qualitative understanding of its complex dynamics rather than quantitative predictions. Low dimension and simple equations enable a parameter-space analysis allowing us to identify preconditions of several asymptotic states and hence fates of humanity and planet. These include a sustainable co-evolution of nature and society, a global collapse and everlasting oscillations. We consider different scenarios corresponding to different socio-cultural stages of human history. The necessity of accounting for the 'human factor' in Earth system models is highlighted by the finding that carbon stocks during the past centuries evolved opposing to what would 'naturally' be expected on a planet without humans. The intensity of biomass use and the contribution of ecosystem services to human well-being are found to be crucial determinants of the asymptotic state in a (pre-industrial) biomass-only scenario without capital accumulation. The capitalistic, fossil-based scenario reveals that trajectories with fundamentally different asymptotic states might still be almost indistinguishable during even a centuries-long transient phase. Given current human population levels, our study also supports the claim that besides reducing the global demand for energy, only the extensive use of renewable energies may pave the way into a sustainable future.},
author = {Nitzbon, Jan and Heitzig, Jobst and Parlitz, Ulrich},
doi = {10.1088/1748-9326/aa7581},
%file = {:home/paul/Masterarbeit/Papers/Nitzbon_2017_Environ._Res._Lett._12_074020.pdf:pdf},
%issn = {17489326},
journal = {Environ. Res. Lett.},
%keywords = {World-Earth modeling,anthropocene,bifurcation analysis,coevolutionary dynamics,energy transformation,global carbon cycle},
number = {7},
title = {{Sustainability, collapse and oscillations in a simple World-Earth model}},
volume = {12},
year = {2017}
}

@article{Geier2019,
%abstract = {Adaptive networks are a versatile approach to model phenomena such as contagion and spreading dynamics, critical transitions and structure formation that emerge from the dynamic coevolution of complex network structure and node states. Adaptive networks have been successfully applied to study and understand phenomena ranging from epidemic spreading, infrastructure, swarm dynamics and opinion formation to the sustainable use of renewable resources. Here, we study critical transitions in contagion dynamics on multilayer adaptive networks with dynamic node states and present an application to the governance of sustainable resource use. We focus on a three-layer adaptive network model, where a polycentric governance network interacts with a social network of resource users which in turn interacts with an ecological network of renewable resources. We uncover that sustainability is favored for slow interaction timescales, large homophilic network adaptation rate (as long it is below the fragmentation threshold) and high taxation rates. Interestingly, we also observe a trade-off between an eco-dictatorship (reduced model with a single governance actor that always taxes unsustainable resource use) and the polycentric governance network of multiple actors. In the latter setup, sustainability is enhanced for low but hindered for high tax rates compared to the eco-dictatorship case. These results highlight mechanisms generating emergent critical transitions in contagion dynamics on multilayer adaptive networks and show how these can be understood and approximated analytically, relevant for understanding complex adaptive systems from various disciplines ranging from physics and epidemiology to sociology and global sustainability science. The paper also provides insights into potential critical intervention points for policy in the form of taxes in the governance of sustainable renewable resource use that can inform more process-detailed social-ecological modeling.},
%archivePrefix = {arXiv},
%arxivId = {1906.08679},
author = {Geier, Fabian and Barfuss, Wolfram and Wiedermann, Marc and Kurths, J{\"{u}}rgen and Donges, Jonathan F.},
doi = {10.1140/epjst/e2019-900120-4},
%eprint = {1906.08679},
%file = {:home/paul/Masterarbeit/Papers/Geier2019_Article_ThePhysicsOfGovernanceNetworks.pdf:pdf},
%issn = {19516401},
journal = {Eur. Phys. J. Special Topics},%European Physical Journal: Special Topics}

@article{DavidTabara2018,
%abstract = {The challenge of meeting the UNFCCC CoP21 goal of keeping global warming ‘well below 2 °C and to pursue efforts towards 1.5 °C' (‘the 2–1.5 °C target') calls for research efforts to better understand the opportunities and constraints for fundamental transformations in global systems dynamics which currently drive the unsustainable and inequitable use of the Earth's resources. To this end, this research reviews and introduces the notion of positive tipping points as emergent properties of systems–including both human capacities and structural conditions — which would allow the fast deployment of evolutionary-like transformative solutions to successfully tackle the present socio-climate quandary. Our research provides a simple procedural synthesis to help identify and coordinate the required agents' capacities to implement transformative solutions aligned with such climate goal in different contexts. Our research shows how to identify the required capacities, conditions and potential policy interventions which could eventually lead to the emergence of positive tipping points in various social–ecological systems to address the 2–1.5 °C policy target. Our insights are based on the participatory downscaling of global Shared Socio-economic Pathways (SSPs) to Europe, the formulation of pathways of solutions within these scenarios and the results from an agent-based economic modelling.},
author = {T{\`{a}}bara, David  J. and Frantzeskaki, Niki and H{\"{o}}lscher, Katharina and Pedde, Simona and Kok, Kasper and Lamperti, Francesco and Christensen, Jens H. and J{\"{a}}ger, Jill and Berry, Pam},
doi = {10.1016/j.cosust.2018.01.012},
%file = {:home/paul/Masterarbeit/Papers/1-s2.0-S1877343517300854-main.pdf:pdf},
%issn = {18773435},
journal = {Curr. Opin Env. Sust.},
pages = {120--129},
title = {{Positive tipping points in a rapidly warming world}},
volume = {31},
year = {2018}
}

@article{Patz2005,
abstract = {The World Health Organisation estimates that the warming and precipitation trends due to anthropogenic climate change of the past 30 years already claim over 150,000 lives annually. Many prevalent human diseases are linked to climate fluctuations, from cardiovascular mortality and respiratory illnesses due to heatwaves, to altered transmission of infectious diseases and malnutrition from crop failures. Uncertainty remains in attributing the expansion or resurgence of diseases to climate change, owing to lack of long-term, high-quality data sets as well as the large influence of socio-economic factors and changes in immunity and drug resistance. Here we review the growing evidence that climate-health relationships pose increasing health risks under future projections of climate change and that the warming trend over recent decades has already contributed to increased morbidity and mortality in many regions of the world. Potentially vulnerable regions include the temperate latitudes, which are projected to warm disproportionately, the regions around the Pacific and Indian oceans that are currently subjected to large rainfall variability due to the El Nĩo/Southern Oscillation sub-Saharan Africa and sprawling cities where the urban heat island effect could intensify extreme climatic events. {\textcopyright} 2005 Nature Publishing Group.},
author = {Patz, Jonathan A. and Campbell-Lendrum, Diarmid and Holloway, Tracey and Foley, Jonathan A.},
doi = {10.1038/nature04188},
%file = {:home/paul/Masterarbeit/Papers/nature04188.pdf:pdf},
%issn = {14764687},
journal = {Nature},
number = {7066},
pages = {310--317},
%pmid = {16292302},
title = {{Impact of regional climate change on human health}},
volume = {438},
year = {2005}
}

@article{Brook2010,
abstract = {In 2004, the first American Heart Association scientific statement on "Air Pollution and Cardiovascular Disease" concluded that exposure to particulate matter (PM) air pollution contributes to cardiovascular morbidity and mortality. In the interim, numerous studies have expanded our understanding of this association and further elucidated the physiological and molecular mechanisms involved. The main objective of this updated American Heart Association scientific statement is to provide a comprehensive review of the new evidence linking PM exposure with cardiovascular disease, with a specific focus on highlighting the clinical implications for researchers and healthcare providers. The writing group also sought to provide expert consensus opinions on many aspects of the current state of science and updated suggestions for areas of future research. On the basis of the findings of this review, several new conclusions were reached, including the following: Exposure to PM <2.5 $\mu$m in diameter (PM2.5) over a few hours to weeks can trigger cardiovascular disease-related mortality and nonfatal events; longer-term exposure (eg, a few years) increases the risk for cardiovascular mortality to an even greater extent than exposures over a few days and reduces life expectancy within more highly exposed segments of the population by several months to a few years; reductions in PM levels are associated with decreases in cardiovascular mortality within a time frame as short as a few years; and many credible pathological mechanisms have been elucidated that lend biological plausibility to these findings. It is the opinion of the writing group that the overall evidence is consistent with a causal relationship between PM2.5 exposure and cardiovascular morbidity and mortality. This body of evidence has grown and been strengthened substantially since the first American Heart Association scientific statement was published. Finally, PM2.5 exposure is deemed a modifiable factor that contributes to cardiovascular morbidity and mortality. {\textcopyright} 2010 American Heart Association, Inc.},
author = {Brook, Robert D. and Rajagopalan, Sanjay and Pope, C. Arden and Brook, Jeffrey R. and Bhatnagar, Aruni and Diez-Roux, Ana V. and Holguin, Fernando and Hong, Yuling and Luepker, Russell V. and Mittleman, Murray A. and Peters, Annette and Siscovick, David and Smith, Sidney C. and Whitsel, Laurie and Kaufman, Joel D.},
doi = {10.1161/CIR.0b013e3181dbece1},
%file = {:home/paul/Masterarbeit/Papers/CIR.0b013e3181dbece1.pdf:pdf},
%issn = {00097322},
journal = {Circulation},
%keywords = {AHA Scientific Statements,Air pollution,Atherosclerosis,Epidemiology,Prevention,Public policy},
number = {21},
pages = {2331--2378},
%pmid = {20458016},
title = {{Particulate matter air pollution and cardiovascular disease: An update to the scientific statement from the american heart association}},
volume = {121},
year = {2010}
}

@article{McMichael2012,
abstract = {Climate change poses threats to human health, safety, and survival via weather extremes and climatic impacts on food yields, fresh water, infectious diseases, conflict, and displacement. Paradoxically, these risks to health are neither widely nor fully recognized. Historical experiences of diverse societies experiencing climatic changes, spanning multicentury to single-year duration, provide insights into population health vulnerability - even though most climatic changes were considerably less than those anticipated this century and beyond. Historical experience indicates the following. (i) Long-term climate changes have often destabilized civilizations, typically via food shortages, consequent hunger, disease, and unrest. (ii) Medium-term climatic adversity has frequently caused similar health, social, and sometimes political consequences. (iii) Infectious disease epidemics have often occurred in association with briefer episodes of temperature shifts, food shortages, impoverishment, and social disruption. (iv) Societies have often learnt to cope (despite hardship for some groups) with recurring shorterterm (decadal to multiyear) regional climatic cycles (e.g., El Ni{\~{n}}o Southern Oscillation) - except when extreme phases occur. (v) The drought-famine-starvation nexus has been the main, recurring, serious threat to health. Warming this century is not only likely to greatly exceed the Holocene's natural multidecadal temperature fluctuations but to occur faster. Along with greater climatic variability, models project an increased geographic range and severity of droughts. Modern societies, although larger, better resourced, and more interconnected than past societies, are less flexible, more infrastructure-dependent, densely populated, and hence are vulnerable. Adverse historical climate-related health experiences underscore the case for abating human-induced climate change.},
author = {McMichael, A. J.},
doi = {10.1073/pnas.1120177109},
%file = {:home/paul/Masterarbeit/Papers/4730.full.pdf:pdf},
%issn = {00278424},
journal = {P.  Natl. Acad Sci. USA},%Proceedings of the National Academy of Sciences of the United States of America}

@article{McMichael2011,
%abstract = {Recent observed changes in Earth's climate, to which humans have contributed substantially, are affecting various health outcomes. These include altered distributions of some infectious disease vectors (ticks at high latitudes, malaria mosquitoes at high altitudes), and an uptrend in extreme weather events and associated deaths, injuries and other health outcomes. Future climate change, if unchecked, will have increasing, mostly adverse, health impacts - both direct and indirect. Climate change will amplify health problems in vulnerable regions, influence infectious disease emergence, affect food yields and nutrition, increase risks of climate-related disasters and impair mental health. The health sector should assist society understand the risks to health and the needed responses. {\textcopyright} 2011 The Association for the Publication of the J. Intern. Med..},
author = {McMichael, A. J. and Lindgren, E.},
doi = {10.1111/j.1365-2796.2011.02415.x},
%file = {:home/paul/Masterarbeit/Papers/j.1365-2796.2011.02415.x.pdf:pdf},
%issn = {09546820},
journal = {J. Intern. Med.},
%keywords = {Cardiovascular diseases,Climate change,Communicable diseases,Health risks,Injuries,Malnutrition},
number = {5},
pages = {401--413},
%pmid = {21682780},
title = {{Climate change: Present and future risks to health, and necessary responses}},
volume = {270},
year = {2011}
}

@article{Zhao1998,
abstract = {Based on 70 interviews with informants who were mostly students during the 1989 Beijing student movement, the author found that the ecology of university campuses in Beijing enclosed a huge number of students in a small area with a unique spatial distribution and regulated their spatial activities. This ecology nurtured many close-knit student networks, as well as directly exposed all Beijing students to a collective action environment when the movement started. These ecological conditions not only sustained a high rate of movement participation but also facilitated the formation of many ecology-dependent strategies of student mobilization, which in turn patterned the dynamics of the movement.},
author = {Zhao, Dingxin},
doi = {10.1086/231399},
%file = {:home/paul/Masterarbeit/Papers/231399.pdf:pdf},
%issn = {00029602},
journal = {Am. J. Sociol},
number = {6},
pages = {1493--1529},
title = {{Ecologies of social movements: Student mobilization during the 1989 prodemocracy movement in Beijing}},
volume = {103},
year = {1998}
}

@article{Gould1993,
author = {Gould, Roger V},
%file = {:home/paul/Masterarbeit/Papers/2095965.pdf:pdf},
journal = {Am. Sociol. Rev.},
number = {2},
pages = {182--196},
title = {{Collective action and network structure}},
volume = {58},
year = {1993}
}

@article{Gould1991,
author = {Gould, Roger V},
%file = {:home/paul/Masterarbeit/Papers/2096251.pdf:pdf},
journal = {Am. Sociol. Rev.},
number = {6},
pages = {716--729},
title = {{Multiple networks and mobilization in the Paris commune, 1871}},
volume = {56},
year = {1991}
}

@article{Centola2005,
author = {Centola, Damon and Willer, Robb and Macy, Michael W},
%file = {:home/paul/Masterarbeit/Papers/427321.pdf:pdf},
journal = {Am. J. Sociol.},
number = {4},
pages = {1009--1040},
title = {{The emperor’s dilemma: A computational model of self-enforcing norms}},
volume = {110},
year = {2005}
}

@article{Centola2007,
abstract = {The strength of weak ties is that they tend to be long—they connect socially distant locations, allowing information to diffuse rapidly. The authors test whether this “strength of weak ties” generalizes from simple to complex contagions. Complex contagions require social affirmation from multiple sources. Examples include the spread of high-risk social movements, avant garde fashions, and unproven technologies. Results show that as adoption thresholds increase, long ties can impede diffusion. Complex contagions depend primarily on the width of the bridges across a network, not just their length.Wide bridges are a characteristic feature of manyspatial networks, which may account in part for the widely observed ten- dency for social movements to diffuse spatially.},
author = {Centola, Damon and Macy, Michael W},
%file = {:home/paul/Masterarbeit/Papers/521848.pdf:pdf},
journal = {Am. J. Sociol.},
number = {3},
pages = {702--734},
title = {{Complex contagions and the weakness of long ties}},
volume = {113},
year = {2007}
}

@article{Hagedorn2019,
title = "The concerns of the young protesters are justified: A statement by Scientists for Future concerning the protests for more climate protection",
journal = "GAIA ",
%parent_itemid = "infobike://oekom/gaia",
%publishercode ="oekom",
year = "2019",
volume = "28",
number = "2",
%publication date ="2019-01-01T00:00:00",
pages = "79-87",
%itemtype = "ARTICLE",
%issn = "0940-5550",
%url = "https://www.ingentaconnect.com/content/oekom/gaia/2019/00000028/00000002/art00004",
doi = "doi:10.14512/gaia.28.2.3",
%keyword = "science-policy interface, Fridays for Future, climate action, youth movement, sustainability crisis, Climate Strike, biodiversity",
author = "Hagedorn, Gregor and Loew, Thomas and Seneviratne, Sonia I. and Lucht, Wolfgang and Beck, Marie-Luise and Hesse, Janina and Knutti, Reto and Quaschning, Volker and Schleimer, Jan-Hendrik and Mattauch, Linus and Breyer, Christian and H{\"u}bener, Heike and Kirchengast, Gottfried and Chodura, Alice and Clausen, Jens and Creutzig, Felix and Darbi, Marianne and Daub, Claus-Heinrich and Ekardt, Felix and G{\"o}pel, Maja and Judith N., Hardt and Hertin, Julia and Hickler, Thomas and K{\"o}hncke, Arnulf and K{\"o}ster, Stephan and Krohmer, Julia and Kromp-Kolb, Helga and Leinfelder, Reinhold and Mederake, Linda and Neuhaus, Michael and Rahmstorf, Stefan and Schmidt, Christine and Schneider, Christoph and Schneider, Gerhard and Seppelt, Ralf and Spindler, Uli and Springmann, Marco and Staab, Katharina and Stocker, Thomas F. and Steininger, Karl and Hirschhausen, Eckart von and Winter, Susanne and Wittau, Martin and Zens, Josef",
%abstract = "In March 2019, German-speaking scientists and scholars calling themselves Scientists for Future, published a statement in support of the youth protesters in Germany, Austria, and Switzerland (Fridays for Future, Klimastreik/Climate Strike), verifying the scientific evidence
%that the youth protestors refer to. In this article, they provide the full text of the statement, including the list of supporting facts (in both English and German) as well as an analysis of the results and impacts of the statement. Furthermore, they reflect on the challenges for scientists
%and scholars who feel a dual responsibility: on the one hand, to remain independent and politically neutral, and, on the other hand, to inform and warn societies of the dangers that lie ahead.",
}

@article{DeMenocal2001,
%abstract = {Modern complex societies exhibit marked resilience to interannual-to-decadal droughts, but cultural responses to multidecadal-to-multicentury droughts can only be addressed by integrating detailed archaeological and paleoclimatic records. Four case studies drawn from New and Old World civilizations document societal responses to prolonged drought, including population dislocations, urban abandonment, and state collapse. Further study of past cultural adaptations to persistent climate change may provide valuable perspective on possible responses of modern societies to future climate change.},
author = {DeMenocal, P. B.},
doi = {10.1126/science.1059827},
%file = {:home/paul/Masterarbeit/Papers/667.full.pdf:pdf},
%issn = {00368075},
journal = {Science},
number = {5517},
pages = {667--673},
title = {{Cultural responses to climate change during the late holocene}},
volume = {292},
year = {2001}
}

@article{Steffen2011a,
author = {Steffen, Will and Persson, {\AA}sa and Deutsch, Lisa and Zalasiewicz, Jan and Williams, Mark and Richardson, Katherine and Crumley, Carole and Crutzen, Paul and Folke, Carl and Gordon, Line and Molina, Mario and Ramanathan, Veerabhadran and Rockstr{\"{o}}m, Johan and Scheffer, Marten and Schellnhuber, Hans Joachim and Svedin, Uno},
doi = {10.1007/s13280-011-0185-x},
%file = {:home/paul/Masterarbeit/Papers/Steffen2011_Article_TheAnthropoceneFromGlobalChang.pdf:pdf},
%issn = {00447447},
journal = {Ambio},
%keywords = {Anthropocence,Earth System,Ecosystem services,Planetary stewardship,Resilience},
number = {7},
pages = {739--761},
%pmid = {22338713},
title = {{The anthropocene: From global change to planetary stewardship}},
volume = {40},
year = {2011}
}

@article{Brown2019,
author = {Brown, Calum and Seo, Bumsuk and Rounsevell, Mark},
doi = {10.5194/esd-2019-24},
%file = {:home/paul/Masterarbeit/Papers/esd-10-809-2019.pdf:pdf},
%issn = {2190-4979},
journal = {Earth Syst. Dynam},% Discussions}

@article{Muller-Hansen2019,
author = {M{\"{u}}ller-Hansen, Finn and Heitzig, Jobst and Donges, Jonathan F and Cardoso, Manoel F and Dalla-Nora, Eloi L and Andrade, Pedro and Kurths, J{\"{u}}rgen and Thonicke, Kirsten},
doi = {https://doi.org/10.1016/j.ecolecon.2018.12.025},
%issn = {0921-8009},
journal = {Ecol. Econ.},
%keywords = {Agent-based modeling,Amazon deforestation,Land-use intensification,Pasture management,Social-ecological systems},
pages = {198--211},
title = {{Can intensification of cattle ranching reduce deforestation in the Amazon? Insights from an agent-based social-ecological model}},
%url = {http://www.sciencedirect.com/science/article/pii/S0921800918304130},
volume = {159},
year = {2019}
}

@article{Beckage2018,
author = {Beckage, Brian and Gross, Louis J. and Lacasse, Katherine and Carr, Eric and Metcalf, Sara S. and Winter, Jonathan M. and Howe, Peter D. and Fefferman, Nina and Franck, Travis and Zia, Asim and Kinzig, Ann and Hoffman, Forrest M.},
doi = {10.1038/s41558-017-0031-7},
%file = {:home/paul/Masterarbeit/Papers/s41558-017-0031-7.pdf:pdf},
%isbn = {4155801700317},
%issn = {17586798},
journal = {Nat. Clim. Change},
number = {1},
pages = {79--84},
%publisher = {Springer US},
title = {{Linking models of human behaviour and climate alters projected climate change}},
url = {http://dx.doi.org/10.1038/s41558-017-0031-7},
volume = {8},
year = {2018}
}

@article{Cullen2000,
    author = {Cullen, H. M. and deMenocal, P. B. and Hemming, S. and Hemming, G. and Brown, F. H. and Guilderson, T. and Sirocko, F.},
    title = "{Climate change and the collapse of the Akkadian empire: Evidence from the deep sea}",
    journal = {Geology},
    volume = {28},
    number = {4},
    pages = {379-382},
    year = {2000},
%    month = {04},
%    issn = {0091-7613},
    doi = {10.1130/0091-7613(2000)28<379:CCATCO>2.0.CO;2},
    url = {https://doi.org/10.1130/0091-7613(2000)28<379:CCATCO>2.0.CO;2},
%    eprint = {https://pubs.geoscienceworld.org/geology/article-pdf/28/4/379/3520669/i0091-7613-28-4-379.pdf},
}

@article{Klose2020,
%archivePrefix = {arXiv},
%arxivId = {1910.12042},
author = {Klose, Ann Kristin and Karle, Volker and Winkelmann, Ricarda and Donges, Jonathan F.},
%eprint = {1910.12042},
%file = {:home/paul/Masterarbeit/Papers/1910.12042.pdf:pdf},
journal = {Roy. Soc. Open Sci.},
%keywords = {critical threshold,domino effect,earth system,eutroph-,hysteresis,tipping point},
pages = {1--18},
title = {{Dynamic emergence of domino effects in systems of interacting tipping elements in ecology and climate}},
%url = {http://arxiv.org/abs/1910.12042},
year = {2020},
volume = {7},
number = {6},
}

@article{Wunderling2020,
author = {Wunderling, Nico and Donges, Jonathan F and Kurths, J{\"{u}}rgen and Winkelmann, Ricarda},
%file = {:home/paul/Masterarbeit/Papers/esd-2020-18.pdf:pdf},
journal = {Earth Syst. Dynam. Discuss.},% Discussions}

\clearpage
\maketitle
\beginsupplement
\section{Derivation of the Piece--Wise Analytical Solution}
For the mean-field approximation we will provide a piece-wise solution in this supplementary document. We repeat the governing equations of the mean-field model here:
\begin{align}
\dot Y &= \frac{1}{\tau}\left(X-Y\right)\,,\label{eq:Y_dot}\\
\dot X &=  (1-X)\tilde{p}^+-X\tilde{p}^-\,.  \label{eq:Xdot}
\end{align}
With
\begin{align}
 \tilde{p}^- &= \alpha\cdot p_\text{dir}^-+\beta \cdot p_{\text{soc}}^-\cdot p_\text{ant}^-\,,\label{eq:stylizedpp2}\\
 &=\alpha H(Y- \gamma)+ \beta \left[1-H(X-\chi)\right]\cdot H(Y+\frac{\theta}{\tau}(X-Y)-\gamma) \,, \label{eq:pm}\\
     \tilde{p}^+ &=\alpha\cdot p_\text{dir}^++\beta \cdot p_{\text{soc}}^+\cdot p_\text{ant}^+\,.\label{eq:stylizedpm2}\\
     &=\alpha \left[1-H(Y-\gamma)\right]+ \beta H(X-\chi)\cdot\left[1-H(Y+\frac{\theta}{\tau}(X-Y)-\gamma)\right] \label{eq:pp}\,,
\end{align}
being the change rates which are constant except for $Y=\gamma$, $X=\chi$ and $Y+\frac{\theta}{\tau}(X-Y)=\gamma$. Thus, three distinct cases must be solved as $\tilde{p}^-$ and $ \tilde{p}^+$ can either by  constant at zero or non-zero, but not both can be zero simultaneously due to one of the $p^\pm_\text{dir}$ being always one.  The three cases are listed and discussed in the following.
\subsection*{Analytical Solution for $\tilde{p}^->0,\;\tilde{p}^+=0$}
\label{sec:case2}
For $\tilde{p}^->0$  and a $\tilde{p}^+=0$ two conditions must be met:
\begin{enumerate}
    \item The pollution $Y$ is above $\gamma$
    \begin{align*}
    Y\geqslant  \gamma\Rightarrow p_\text{dir}^+=0,\,p_\text{dir}^->0\,.
    \end{align*}
    \item  The share of polluting agents $X$ is below $\chi$ or the anticipated pollution is above $\gamma$ 
     \begin{align*}
    (X<\chi)\;\vee\; (Y +\theta \dot Y \geqslant \gamma)\Rightarrow p_{\text{soc}}^+\cdot p_\text{ant}^+=0\,.
    \end{align*}

\end{enumerate}
Note that $\tilde p^-$ can still be $\tilde p^-=\alpha$ or $\tilde p^-=\alpha+\beta$, depending on  $p_{\text{soc}}^-\cdot p_\text{ant}^-$ being zero or one.
Combining the two above conditions yields
\begin{align}
\left[(Y \geqslant \gamma) \;\wedge\; \left((X<\chi)\;\vee\; (Y +\theta \dot Y \geqslant \gamma)\right)\right]
    \Rightarrow\  \tilde{p}^- >0,\; \tilde{p}^+ = 0 \,.\label{eq:p+1p-0}
\end{align}
Plugging Eq.\;(\ref{eq:p+1p-0}) into Eq.\;(\ref{eq:Xdot}) allows us to solve for $X$ using an exponential ansatz
\begin{align}
\dot X(t) &= -X\tilde{p}^-\,,\\
    \Rightarrow X(t) &= X_0\ex{-\tilde{p}^-t}\,.\label{eq:case2X}
 \end{align}
 The initial condition is given by $X_0=X(t=0)$.
 The pollution can then be obtained by plugging Eq.\;(\ref{eq:case2X}) into Eq.\;(\ref{eq:Y_dot})  and applying the method of variation of constants
 \begin{align}
     Y(t) &= \begin{cases}
     \frac{X_0}{1-\tilde{p}^-\tau}\ex{-\tilde{p}^-t} +\left(Y_0-\frac{X_0}{1-\tilde{p}^-\tau}\right)\ex{-\frac{t}{\tau}},\text{ if }\tilde{p}^-\neq \frac{1}{\tau}\vspace{0.5cm}\\
     X_0\frac{t}{\tau}\ex{-\frac{t}{\tau}}+Y_0\ex{-\frac{t}{\tau}},\text{ if }\tilde{p}^-=\frac{1}{\tau}
     \end{cases} \label{eq:case2Y}\,.
\end{align}
The initial condition is given by $Y_0=Y(t=0)$.
Eqs.~(\ref{eq:case2X},\ref{eq:case2Y}) lead to a convergence of the share of polluting agents  and the pollution to zero in the long-term behaviour
\begin{align}
\lim_{t\rightarrow\infty}X(t)&\rightarrow 0\,,\\
\lim_{t\rightarrow\infty}Y(t)&\rightarrow 0\,.
\end{align} 
However, reaching zero pollution $Y$ should eventually lead to $Y$ being below $\gamma$ which violates condition 1 for this case. Consequently, this case can only be a transient case.
\subsection*{Analytical Solution for $\tilde{p}^-=0,\;\tilde{p}^+>0$}
Having $\tilde{p}^-=0$ and $\tilde{p}^+>0$ can be achieved under the two following conditions:
\begin{enumerate}
    \item The pollution $Y$ is below $\gamma$
    \begin{align*}
    Y<\gamma\Rightarrow p_\text{dir}^-=0,\,p_\text{dir}>0.
\end{align*}     
    \item The share of polluting agents $X$ is above $\chi$ or the anticipated pollution is below $\gamma$
    \begin{align*}
    (X \geqslant\chi) \;\vee\; (Y+\theta \dot Y < \gamma)\Rightarrow p_{\text{soc}}^-\cdot p_\text{ant}^-=0\,.
\end{align*}      
\end{enumerate}
Here, $\tilde{p}^+$ can still be $\tilde{p}^+=\alpha$ or $\tilde{p}^+=\alpha+\beta$, depending on $p_{\text{soc}}^+\cdot p_\text{ant}^+$ being zero or one.
In summary,
\begin{align}
\left[(Y < \gamma) \;\wedge\; \left((X\geqslant\gamma) \;\vee\; (Y +\theta \dot Y < \gamma)\right) \right]
    \Rightarrow\   \tilde{p}^- =0,\; \tilde{p}^+ > 0\,. \label{eq:p+0p-1}
\end{align}
Again, an exponential ansatz can be used to obtain $X$ when plugging Eq.\;(\ref{eq:p+0p-1}) into Eq.\;(\ref{eq:Xdot}) 
\begin{align}
\dot X(t)&= (1-X)\tilde{p}^+\,,\\
    \Rightarrow X(t) &= 1-X_0\ex{-\tilde{p}^+t}\,.\label{eq:case3X}
\end{align} 
Plugging Eq.\;(\ref{eq:case3X}) into Eq.\;(\ref{eq:Y_dot}) and using the method of variation of constants allows to solve for the pollution 
\begin{align}
     Y(t) &= \begin{cases}
     1-\frac{X_0}{1-\tilde{p}^+\tau}\ex{-\tilde{p}^+t} +\left(Y_0+\frac{X_0}{1-\tilde{p}^+\tau}-1\right)\ex{-\frac{t}{\tau}},\text{ if }\tilde{p}^+\neq \frac{1}{\tau}\vspace{0.5cm}\\
     1-X_0t\ex{-\frac{t}{\tau}}+(Y_0-1)\ex{-\frac{t}{\tau}},\text{ if }\tilde{p}^+=\frac{1}{\tau}
     \end{cases} \label{eq:case3Y}\,.
\end{align}
Eqs.~(\ref{eq:case3X},\ref{eq:case3Y}) result in convergence of both, the share of polluting agents $X$ and the pollution $Y$ to one
\begin{align}
\lim_{t\rightarrow\infty}X(t)&\rightarrow 1\,,\\
\lim_{t\rightarrow\infty}Y(t)&\rightarrow 1\,.
\end{align}
However, $Y$ approaching one will  lead to $Y\geqslant \gamma$ which violates the first condition for this case. Consequently, this case can only be a transient case as well.  
The fixed points of this case are on the opposite side of the valid range for $Y$ and $X$ compared to previous case. The cases $\tilde{p}^->0,\ \tilde{p}^+=0$ and  $\tilde{p}^-=0,\ \tilde{p}^+>0$ are therefore contrasting cases which originate from the symmetric formulation of the model. 
\subsection*{Analytical Solution for $\tilde{p}^->0,\;\tilde{p}^+>0$}
\label{sec:case4}
There are only two ways for both change  rates to become positive, $\tilde{p}^->0,\;\tilde{p}^+>0$:
\begin{enumerate}
    \item The pollution $Y$ is above $\gamma$, the share of polluting agents is above $\chi$ and the anticipated pollution is below $\gamma$
\begin{align*}
 (Y \geqslant \gamma) \;\wedge\; (X\geqslant\chi) \;\wedge\; (Y +\theta \dot Y < \gamma)\Rightarrow p_\text{dir}^-,~  p_{\text{soc}}^+\cdot p_\text{ant}^+ >0 \,.
\end{align*}     
    \item The pollution $Y$ is below $\gamma$, the share of polluting agents $X$ is above $\chi$ and the anticipated pollution is above $\gamma$ 
\begin{align*}
  (Y <\gamma)\;\wedge\; (X < \chi )\;\wedge\; (Y+\theta \dot Y \geqslant \gamma) \Rightarrow p_\text{dir}^+,~  p_{\text{soc}}^-\cdot p_\text{ant}^- >0  \,.
\end{align*}    
\end{enumerate}
If either of these conditions is met, it yields
\begin{align}
\begin{split}
&\left((Y \geqslant \gamma) \;\wedge\; (X\geqslant\chi) \;\wedge\; (Y +\theta \dot Y < \gamma)\right) \;\vee\;\\
&\left((Y < \gamma)\;\wedge\; (X < \chi )\;\wedge\; (Y+\theta \dot Y \geqslant \gamma)\right)\,,\label{eq:cond4}
\end{split}\\
    &\Rightarrow \  \tilde{p}^- >0,\; \tilde{p}^+ > 0 \,.\label{eq:p+1p-1}
\end{align}
Plugging Eq.\;\eqref{eq:p+1p-1} into Eq.\;\eqref{eq:Xdot} and using an exponential ansatz allows us to solve for $X$
\begin{align}
\dot X(t) &=\tilde{p}^+(1-X)-\tilde{p}^-X\,,\\
    \Rightarrow X(t) &= q^++\left(X_0-q^+\right)\ex{-st}\,.\label{eq:case4X}  
\end{align}
Here, we substituted
\begin{align}
s &= \tilde{p}^- +\tilde{p}^+\,,\\
q^+&=\frac{\tilde{p}^+}{s}=\frac{\tilde{p}^+}{\tilde{p}^- +\tilde{p}^+} \,.
\end{align}
If we plug Eq.\;\eqref{eq:case4X} into  Eq.\;\eqref{eq:Y_dot} and use the method of variations of constants again, we can solve for $Y$
\begin{align}
     Y(t) &= \begin{cases}
        &q^++\frac{X_0 -q^+}{1-s\tau}\ex{-st}+ 
     \left(Y_0-q^+-\frac{X_0 -q^+}{1-s\tau}\right)\ex{-\frac{t}{\tau}},\text{ if }s\neq \frac{1}{\tau} 
    \vspace{.5cm}\\
         &q^++(X_0-q^+)\frac{t}{\tau}\ex{-\frac{t}{\tau}} +(Y_0-q^+)\ex{-\frac{t}{\tau}},\text{if } s=\frac{1}{\tau}
     \end{cases} \label{eq:case4Y}\,.
\end{align}
Both change rates being positive causes the share of polluting agents and the pollution to approach a fixed point
\begin{align}
\lim_{t\rightarrow\infty}X(t)&= q^+\,,\\
\lim_{t\rightarrow\infty}Y(t)&= q^+\,.
\end{align}
 As approaching a fixed point would eventually lead to vanishing change rates of the pollution $\dot Y =0$, the anticipated pollution $Y+\theta\dot Y$ becomes the actual pollution $Y$
 \begin{align}
 \lim_{t\rightarrow\infty}Y+\theta\dot Y -Y=0\,.\label{eq:contra1}
 \end{align}
 However, to meet either of the conditions for this case the anticipated pollution $Y+\theta \dot Y$ and the actual pollution $Y$ have to be at  opposite sides of the pollution threshold $\gamma$, compare Eq.\;\eqref{eq:cond4} which is violated by Eq.\;\eqref{eq:contra1}. Thus, before the fixed points are reached the case of the analytical solution has to change. Therefore, this case can only be a transient one, too. 
\\\\
Every case on its own would approach a fixed point for $Y$ and $X$, however, the fixed points are always in contradiction with the necessary condition of $Y$, $\dot Y$ and $X$ compared to the thresholds $\gamma$ and $\chi$. To tackle this problem and find a full analytical solution one would have to evaluate the times when to switch between different cases of the piece-wise solution, i.e., when a threshold is passed. However, this is generally not possible due to the discontinuous nature of $\dot X$. Therefore, we  asses the equations numerically through integrating for discrete time steps. In this case the probability to hit a boundary goes to zero.
In the following section we will try to give an approximated solution for the boundary case of $\theta\rightarrow\infty$.

\section{Stability in the Mean-Field Approximation}

\begin{figure}
    \centering
    \includegraphics[trim=200 50 280 0,width=0.5\textwidth]{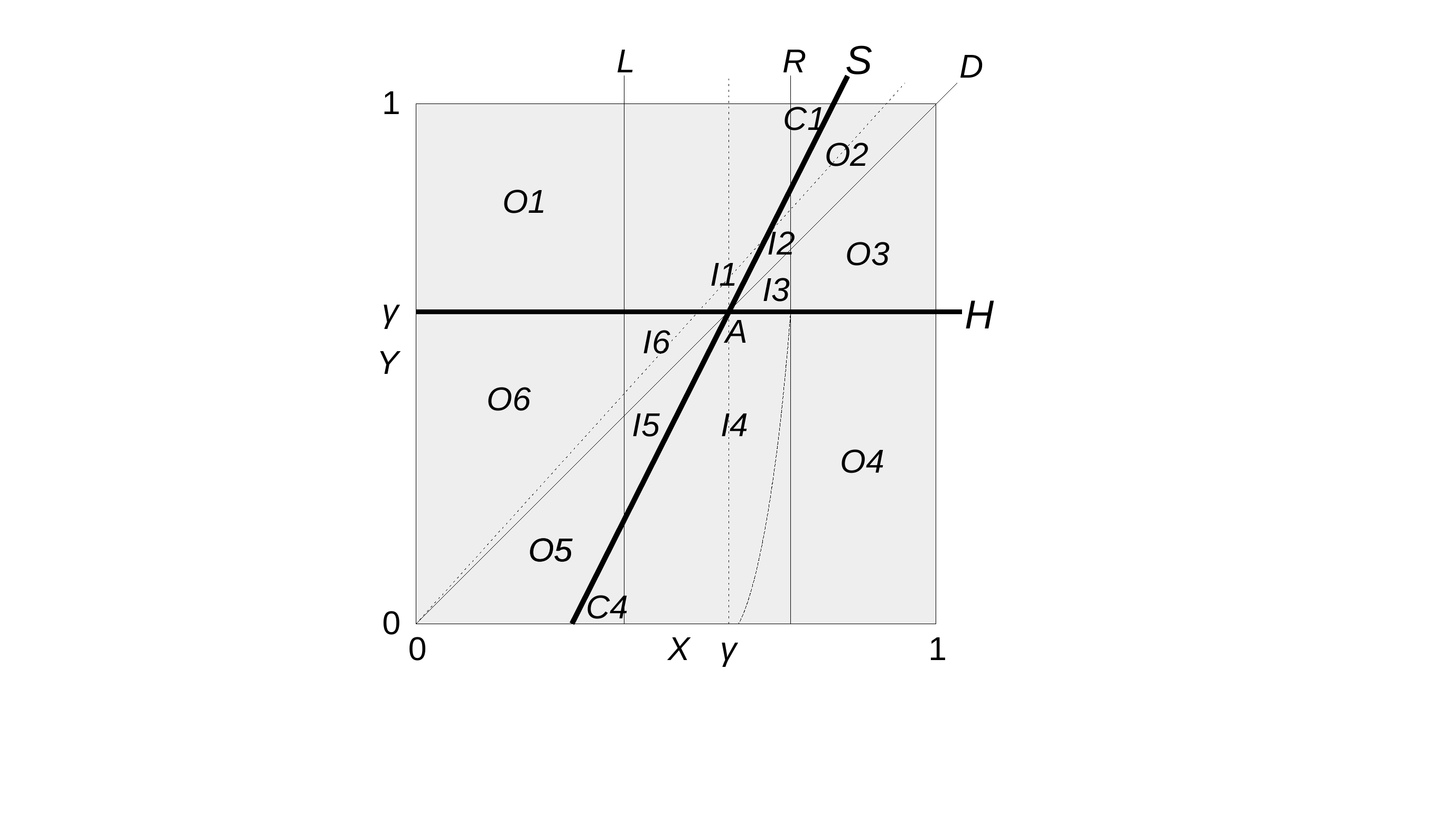}
    \caption{Division of macroscopic state space into regions relevant for analysing the dynamics. Thick lines mark the locus of discontinuities in the macroscopic approximation.}
    \label{fig:division}
\end{figure}

From Eqs.~(\ref{eq:Y_dot},\ref{eq:Xdot},\ref{eq:pm},\ref{eq:pp})

we see that the unit square $U = [0,1]\times[0,1]$ of possible macro-states $(X,Y)$ can be divided by two lines, a horizontal line $H$ given by $Y = \gamma$, and a slanted line $S$ given by $Y_L = Y + \frac{\theta}{\tau} (X-Y) = \gamma$, into four regions (Fig.~\ref{fig:division}). Whenever the system's trajectory hits one of those lines $\dot X$ becomes discontinuous. $H$ and $S$ intersect at $X=Y=\gamma$ and have slopes of $0$ and $s = \frac{\theta}{\theta - \tau}$, respectively. We only analyze the case of $\theta > \tau$ here so that $s > 0$.

In the two top regions, $Y > \gamma$ so that $\tilde p^-_\text{dir}=1$ but $\tilde p^+\text{dir}=0$. In the two bottom regions, $Y < \gamma$ so that $\tilde p^+\text{dir}=1$ but $\tilde p^\text{dir}=0$.
In the two left regions, $Y_L < \gamma$ so that $p_{\text{soc}}^-\cdot p_\text{ant}^-=0$, and $p_{\text{soc}}^1\cdot p_\text{ant}^+=1$  iff $X\ge \chi$.
In the two right regions, $Y_L > \gamma$ so that $p_{\text{soc}}^1\cdot p_\text{ant}^+=0$, and $p_{\text{soc}}^-\cdot p_\text{ant}^-=1$ iff $X<\chi$.

Let us now assume that $\chi < \gamma < \beta/(\alpha + \beta)$ as in our simulations, and further divide the four regions by the two vertical lines $L$ given by $X = \chi$ and $R$ given by $X = \beta/(\alpha + \beta)$ (also marked in Fig.~\ref{fig:division}). We will see that these divisions are relevant for the signs of $\dot X$ and $\dot Y$. 
If $\theta > \tau\gamma/\chi$, $L$ intersects $S$ at $Y = \frac{\theta\chi - \tau\gamma}{\theta - \tau} > 0$. 
If $\theta > \frac{1-\gamma}{1-\beta/(\alpha+\beta)}$,
$R$ intersects $S$ at $Y = \frac{\theta\beta/(\alpha+\beta) - \tau\gamma}{\theta - \tau} < 1$.

Finally, we divide all regions also by the main diagonal $D$ given by $X=Y$. Overall, $U$ is then divided into at least 12 and at most 14 regions --- six ``inner'' regions I1--6, six ``outer'' regions O1--6, and two ``conditional'' regions C1,4 --- as depicted in Fig.~\ref{fig:division}, depending on whether S intersects L and/or R.

We will now show that the central intersection point $A$ given by $X=Y=\gamma$ is a stable attractor in the sense that there is an open neighbourhood of it so that trajectories starting in there will converge to $A$.

Inside region I2, both $\dot X,\dot Y < 0$, hence the system can leave I2 only towards I3 or I1 or stay inside I2. Since $|\dot X|$ is bounded from below by $\gamma\alpha > 0$, it cannot stay inside I2. The slope of the gradient is $g = \dot Y/\dot X = (Y-X)/X\tau\alpha$. Since close to $D$ (below the dotted diagonal in Fig.~\ref{fig:division}), we have $Y < (1+\tau\alpha)X$ and thus $g < 1$, it cannot leave towards I3 across $D$ since the latter has the steeper slope of 1. Hence, the system must leave towards I1, i.e., reach $S$ in finite time.
On $S$, the macroscopic dynamics has a discontinuity and the approximation of the microbehavior is bad. For the micromodel, the above reasoning implies that with a probability approaching one as $N\to\infty$ it will enter that part of I1 where $X>\gamma$ (to the right of the dotted vertical line) in finite time. 
Once there, still $\dot Y < 0$ but now $\dot X > 0$, hence the system can only stay there or go back to I2. Since $\dot Y < 0$ even on the $H$ boundary of I1, it cannot stay in I1. Hence it goes back to I2. Overall, the system switches fast between I2 and I1, all the time having $\dot Y < 0$, hence it must converge to the intersection of $S$ and $H$.

From the rest of I1 (left of the dotted vertical), the trajectory can only go to I2 or I6.
In I6, we have $\dot X > 0 > \dot Y$ and $\dot X > (1-\gamma)(\alpha+\beta) > 0$, hence it must go to I5.
In I5, both $\dot X,\dot Y > 0$, and it cannot stay there since $\dot X > (1-\gamma)(\alpha+\beta) > 0$, hence it must either go back to I6 or to the part of I4 where $X < \gamma$.
If it would go back to I6 every time, it would eventually converge to $A$ since $\dot X > (1-\gamma)(\alpha+\beta) > 0$ throughout. 

In I4, also both $\dot X,\dot Y > 0$, hence the system can either go to I3 or to O4.
The slope of the gradient is $\frac{X-Y}{(1-X)\alpha} > \frac{X-\gamma}{(1-\chi)\alpha} =: f(X)$. Assume the system goes to O4. Then it must start at a point in I4 to the right of the parabola with $\partial Y/\partial X  = f(X)$ and $Y(X=\beta/(\alpha+\beta)) = \gamma$ (sketched as a dashed line in the figure). One can  show that when $\gamma^2 - 2\gamma((1-\chi)\alpha + \beta/(\alpha+\beta)) + \beta^2/(\alpha+\beta)^2$, which is fulfilled in our case, then the whole segment given by $X=\gamma>Y$ (dotted line at $\gamma$ in Fig.~\ref{fig:division}) lies to the left of that parabola, so the system is left of the parabola when coming from I5 and also when starting in a small neighborhood of $A$. Hence the system must go to I3 when in I4.

Finally, in I3, it must go to I2 in finite time because of $\dot X < 0 < \dot Y$ and $|\dot X| > \gamma\alpha > 0$.
Overall, we see that wherever the system starts close to $A$, it will end up converging to $A$ either along the boundary of I2 and I1 or along the boundary of I6 and I5.

\section{Boundary Case: $\theta \rightarrow \infty$}
\label{sec:spec_case} 
For the extreme case of infinite anticipation, i.e.,~$\theta \rightarrow \infty$ we will estimate when deviation of the anticipated pollution from the pollution threshold  $Y+\frac{\theta}{\tau}(X-Y)-\gamma=0$ as it determines the value of $p^\pm_\text{ant}$. With the anticipation time $\theta$ approaching infinity 
this deviation is dominated by $\frac{\theta}{\tau}(X-Y)$ .
With this we can estimate the first switching time $t_\text{switch}$ between two cases of the analytical solution if  the change of $p_\text{ant}^-$ or  $p_\text{ant}^+$ is responsible for the switching. We will present this for the  example of a change due to $p_\text{ant}^+$ changing its value.

We assume  $Y\geqslant \gamma$ so that $\tilde{p}^-_\text{dir}=1$ and $\tilde{p}^+_\text{dir}=0$. 
Additionally, we assume $X(t< t_\text{switch})\geqslant\chi$ so that $p_\text{soc}^+=1$. We start with initial conditions of $X_0>Y_0$ so that $\dot Y >0$ and $\tilde{p}^+_\text{ant}=0$. Under this condition a switching $p_\text{ant}^+$ would lead to change of $\tilde{p}^+$ from  $\tilde{p}^+=0$ to $\tilde{p}^+= \beta$.   Due to the anticipated pollution being dominated by $\theta \dot Y$,  $p_\text{ant}^+$ should become $p_\text{ant}^+=1$ as soon as  the pollution starts decreasing, i.e., $\dot Y<0$. Firstly, we evaluate the time derivative of the pollution $Y$ from Eq.\;(\ref{eq:case2Y})
\begin{align}
    \dot Y (t) &=\frac{-\tilde{p}^-X_0}{1-\tilde{p}^-\tau}\ex{-\tilde{p}^-t}-\frac{1}{\tau}\left(Y_0-\frac{X_0}{1-\tilde{p}^-\tau}\right) \ex{-\frac{t}{\tau}}\,. \label{eq:calc_t_switch} 
\end{align}
We constrain ourselves for the following derivation to $\tilde{p}^-\neq \frac{1}{\tau}$.
Eq.\;(\ref{eq:calc_t_switch}) can  be equated to  zero and solved for $t$ to obtain the switching time $t_\text{switch}$
\begin{align}
    t_\text{switch}&= \frac{\tau}{1-\tilde{p}^-\tau}\ln{\left(1+\frac{1-\tilde{p}^-\tau}{\tilde{p}^-\tau X_0}(X_0-Y_0)\right)}\,. \label{eq:switching_time} 
\end{align}
We omit explicitly showing that $t_\text{switch}$ exists and is always positive, however this is straight forward possibly by calculating the sign of the logarithm depending on the value of $\tilde{p}^-\tau$ and comparing it to the prefactor. 
 
Next, we investigate the dynamics shortly after $t_\text{switch}$. 
We calculate $t_\text{switch}$ under the assumption that the sign of the derivative of the pollution would change. Under the given assumptions and with the anticipated pollution being dominated by  $\frac{\theta}{\tau}(X-Y)$ a switching of $p_\text{ant}^+=0$ to $p_\text{ant}^+=1$ occurs at $t_\text{switch}$. Consequently, the change rate to become polluting changes from $\tilde{p}^+=0$ to $ \tilde{p}^+=\beta$ and the analytical solutions as in Eq.\;(\ref{eq:case4X},\ref{eq:case4Y}) have to be used. 

We now analyse the resulting trajectory further.
Therefore, we develop Eq.\;(\ref{eq:case4Y}) into a Taylor series around $\tilde{t}=0$ with $\tilde{t}=t-t_\text{switch}$. We define $X_1=X(t_\text{switch})$ as the share of polluting agents  and $Y_1=Y(t_\text{switch})$ as the pollution at the switching time   calculated in the previous step. This yields
\begin{align}
\begin{split}
    Y(t) &= Y_1 + \underbrace{(X_1 - Y_1)}_{=0} \tilde{t} \vspace{.5cm}\\
    &+ \frac{1}{2}\cdot\frac{1}{\tau}\left(\frac{s^2\tau(X_1-q^+)}{1-s\tau} +\frac{1}{\tau}\left(Y_1-q^+-\frac{X_1-q^+}{1-s\tau}\right)\right){\tilde{t}}^2 
    +O\left(\tilde{t}\right)^3\,.
\end{split}\label{eq:taylor}
\end{align}
 The first order in $\tilde{t}$ vanishes due to Eq.\;\eqref{eq:Y_dot}  as we calculated  the switching time under the assumption $\dot Y = 0$. We neglect the contributions of orders higher than second order and only investigate non-constant terms as we are interested in the direction of the pollution dynamics.
The only remaining non-constant term is of order two. The coefficient before  $\tilde{t}^2$ is  given by the curvature of the pollution at the switching time $\ddot Y$ which determines the sign of this term 
\begin{align}
  \ddot{Y}(\tilde{t}=0) &= \frac{1}{\tau}\left(\frac{s^2\tau(X_1-q^+)}{1-s\tau} +\frac{1}{\tau}\left(Y_1-q^+-\frac{X_1-q^+}{1-s\tau}\right)\right)\,,\\
  &=\frac{1}{\tau^2}\left(Y_1-X_1(s\tau+1)+s\tau q-\right)\,.
  \end{align}
Here we can use again that $X_1=Y_1$ at the switching time, leading to
\begin{align}
 \ddot{Y}(\tilde{t}=0)&=
\frac{s}{\tau}(q^+-Y_1)\,,\\
&=\frac{s}{\tau}\left(\frac{\tilde{p}^+}{\tilde{p}^++\tilde{p}^-}-Y_1\right)\,,\\
\Rightarrow \ddot{Y}(\tilde{t}=0)&>0 \Leftrightarrow   \frac{\tilde{p}^+}{\tilde{p}^++\tilde{p}^-}>Y_1
  \,.\label{eq:spec_cond}
\end{align}
We found that $\ddot{Y}(\tilde{t}=0)$ is larger than zero in the vicinity of the switching time for the condition in Eq.\;(\ref{eq:spec_cond}). This condition holds almost always  if $\tilde{p}^+\gg \tilde{p}^-$ ($\beta \gg \alpha $) and consequently  the pollution would grow. Due to a large $\theta$ and thus an anticipated pollution  dominated by  $\frac{\theta}{\tau}(X-Y)$, a growing $Y$  will lead to an anticipated pollution above $\gamma$. However, this contradicts  the  conditions for $p_\text{ant}^+=1$ and therefore also  $\tilde{p}^+=\beta$. Consequently, the case of the analytical solution  $\tilde{p}^+>0,\  \tilde{p}^->0$ would immediately switch back to  $\tilde{p}^->0,\  \tilde{p}^+=0$. This leads to the same conditions as right before $t_\text{switch}$. An infinitely fast alternation between the two cases occurs and the pollution level $Y$ becomes fixed at $Y_1$.
Choosing the farsightedness way larger than the vulnerability ($\beta \gg \alpha$) can induce this case. Here, the motivation to become non-polluting due to neighbours and anticipated pollution levels outweighs the motivation to become polluting due to the actual pollution being higher than $\gamma$.

The derivation presented here is only valid under the  assumptions $Y\geqslant \gamma$, $X(t< t_\text{switch})\geqslant\chi$ and $Y_0>X_0$. However, it is possible to derive it under for $Y< \gamma$, $X(t< t_\text{switch})<\chi$ and $Y_0<X_0$ or  calculate the trajectories until the needed conditions are met using a numerical implementation. In general, this derivation makes it clear that a numerical implementation of the mean-field approximation is necessary to address the non-smoothness of the differential equations. However, it also shows that the trajectories could become fixed for high anticipation times at many pollution levels given by $Y(t=t_\text{switch})$. 

\end{document}